\documentclass[prd,superscriptaddress,a4paper,showpacs,showkeys,10pt,nofootinbib]{revtex4}
\usepackage{graphicx}
\usepackage{graphics}
\usepackage{epsfig}
\usepackage{dcolumn}
\usepackage{bm}
\usepackage{multirow}
\usepackage{tabularx}
\usepackage{hyperref}
\usepackage{commath}
\usepackage{epstopdf}
\usepackage[T1]{fontenc}
\usepackage{geometry}
\usepackage{color}

\def\be{\begin{equation}}
\def\ee{\end{equation}}

\providecommand{\ee}{e$^+$e$^-$}

\providecommand{\tabularnewline}{\\}

\makeatother

\newcommand{\pom}{\tt I\! P}

\begin{document}
% 
% \begin{flushright}
% LU TP 16-XX\\
% October 2016
% \vskip1cm
% \end{flushright}
% 
\title{Exclusive and diffractive $\gamma \gamma$ production in $PbPb$ collisions at the LHC, HE -- LHC and FCC}

%%%%%%%%%%%%%%%%%%%%%%%%%%%%%%%%%%%%%%%%%%%%%%%%%%%%%%%%%%%%%%%%%%

%%%%%%%%%%%%%%%%%%%%%%%%%%%%%%%%%%%%%%%%%%%%%%%%%%%%%%%%%%%%%%%%%%

\author{R. O. Coelho}

\email[]{coelho72@gmail.com}

\affiliation{Instituto de F\'{\i}sica e Matem\'atica, Universidade Federal de
Pelotas (UFPel),\\
Caixa Postal 354, CEP 96010-090, Pelotas, RS, Brazil}

\author{V. P. Gon\c calves}

\email[]{barros@ufpel.edu.br}

\affiliation{Instituto de F\'{\i}sica e Matem\'atica, Universidade Federal de
Pelotas (UFPel),\\
Caixa Postal 354, CEP 96010-090, Pelotas, RS, Brazil}

\author{D. E. Martins}

\email[]{dan.ernani@gmail.com}

\affiliation{Instituto de F\'isica, Universidade Federal do Rio de Janeiro (UFRJ), 
Caixa Postal 68528, CEP 21941-972, Rio de Janeiro, RJ, Brazil}

\author{M. Rangel}

\email[]{rangel@if.ufrj.br}

\affiliation{Instituto de F\'isica, Universidade Federal do Rio de Janeiro (UFRJ), 
Caixa Postal 68528, CEP 21941-972, Rio de Janeiro, RJ, Brazil}

%%%%%%%%%%%%%%%%%%%%%%%%%%%%%%%%%%%%%%%%%%%%%%%%%%%%%%%%%%%%%%%%%%

%%%%%%%%%%%%%%%%%%%%%%%%%%%%%%%%%%%%%%%%%%%%%%%%%%%%%%%%%%%%%%%%%%

\begin{abstract}
In this paper we present a detailed analysis of the contribution of the Light -- by -- Light (LbL), Durham and double diffractive processes for the diphoton production in ultraperipheral $PbPb$ collisions at the  Large Hadron Collider (LHC), High -- Energy LHC (HE -- LHC) and Future Circular Collider (FCC).
The acceptance of the central and forward LHC detectors is taken into account and predictions for the invariant mass, rapidity, transverse momentum and acoplanarity distributions are presented. Our results indicate that the contribution of the Durham and double diffractive processes can be strongly suppressed by the exclusivity cuts, which will allow to perform a precise analysis of the LbL scattering, as well the search of beyond Standard Model physics in this final state.
\end{abstract}

%%%%%%%%%%%%%%%%%%%%%%%%%%%%%%%%%%%%%%%%%%%%%%%%%%%%%%%%%%%%%%%%%%



\maketitle

\section{Introduction}

\begin{figure}
\begin{tabular}{cc}
\hspace{-1cm}
{\psfig{figure=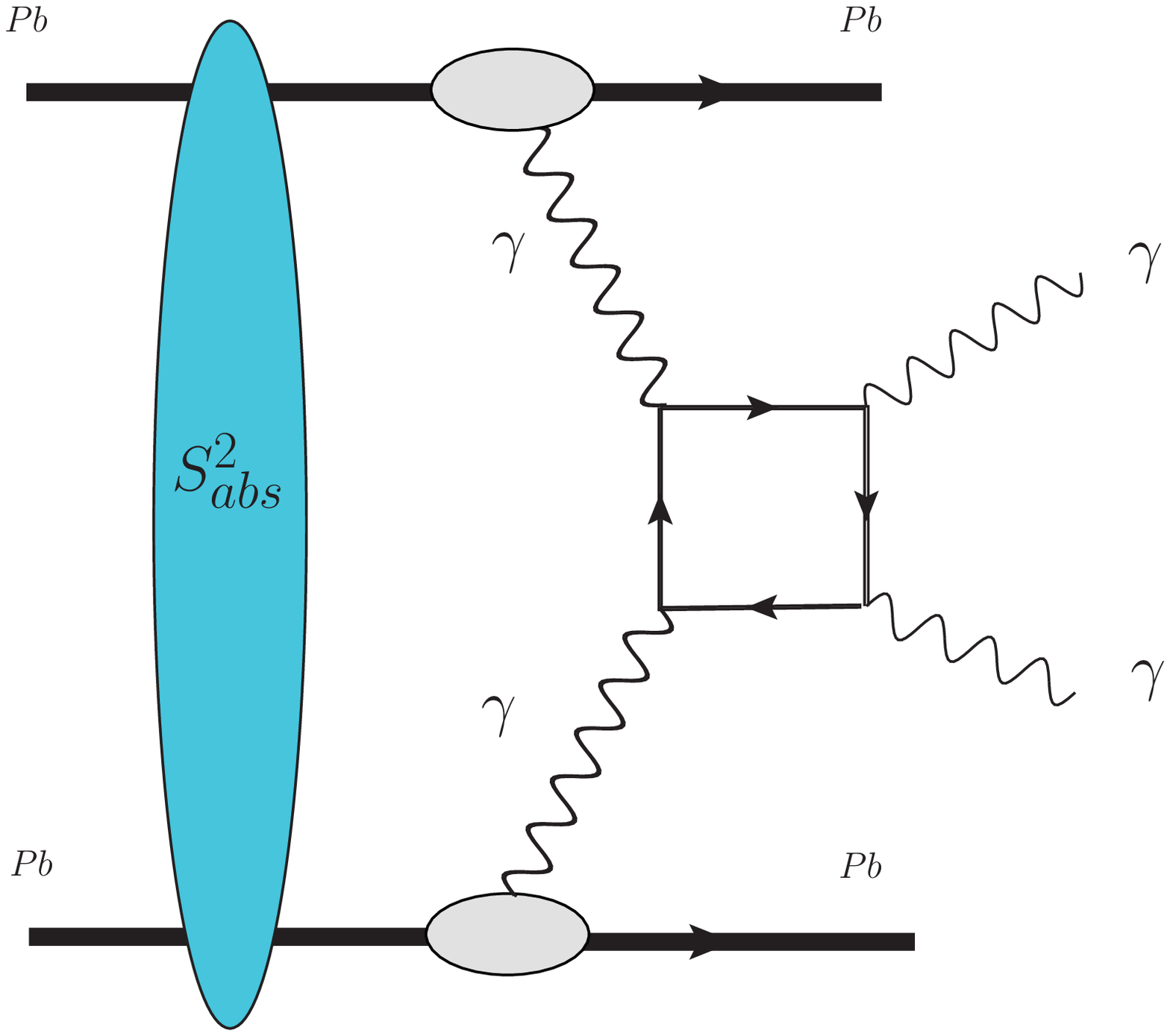,width=5.2cm}} & 
{\psfig{figure=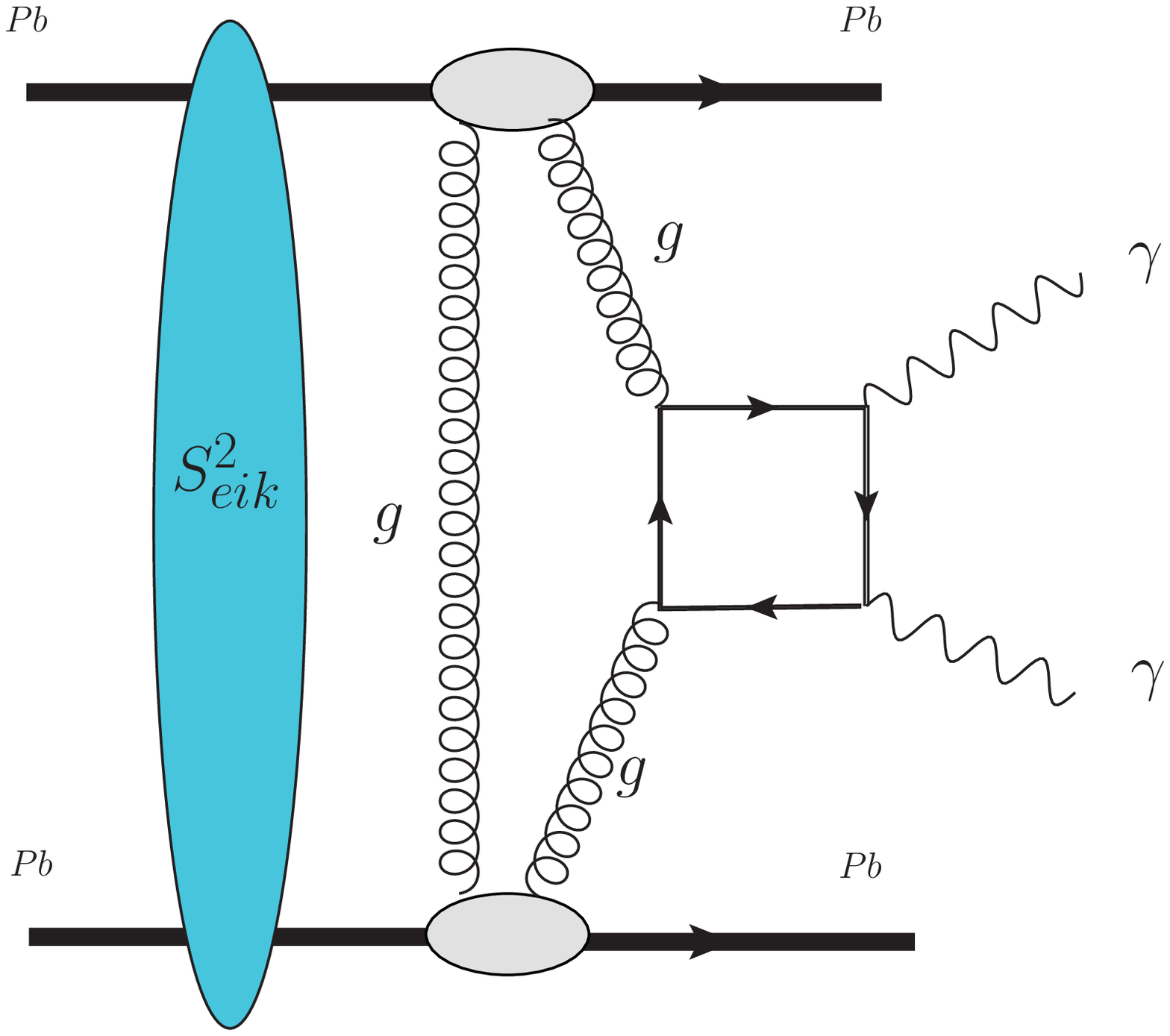,width=5.2cm}} \\ 
(a) & (b) \\ 
\hspace{-1cm}
{\psfig{figure=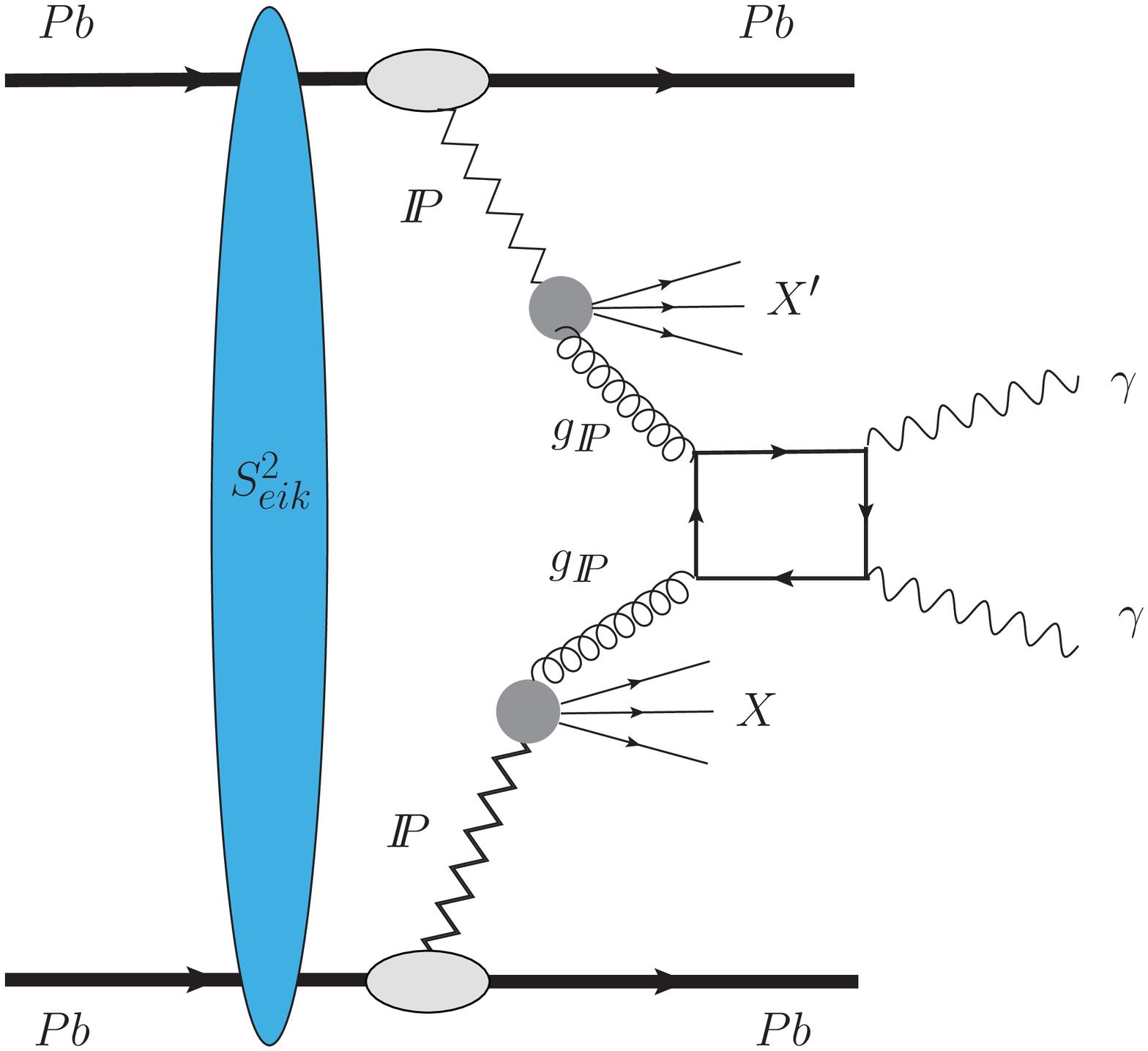,width=5.2cm}}&
{\psfig{figure=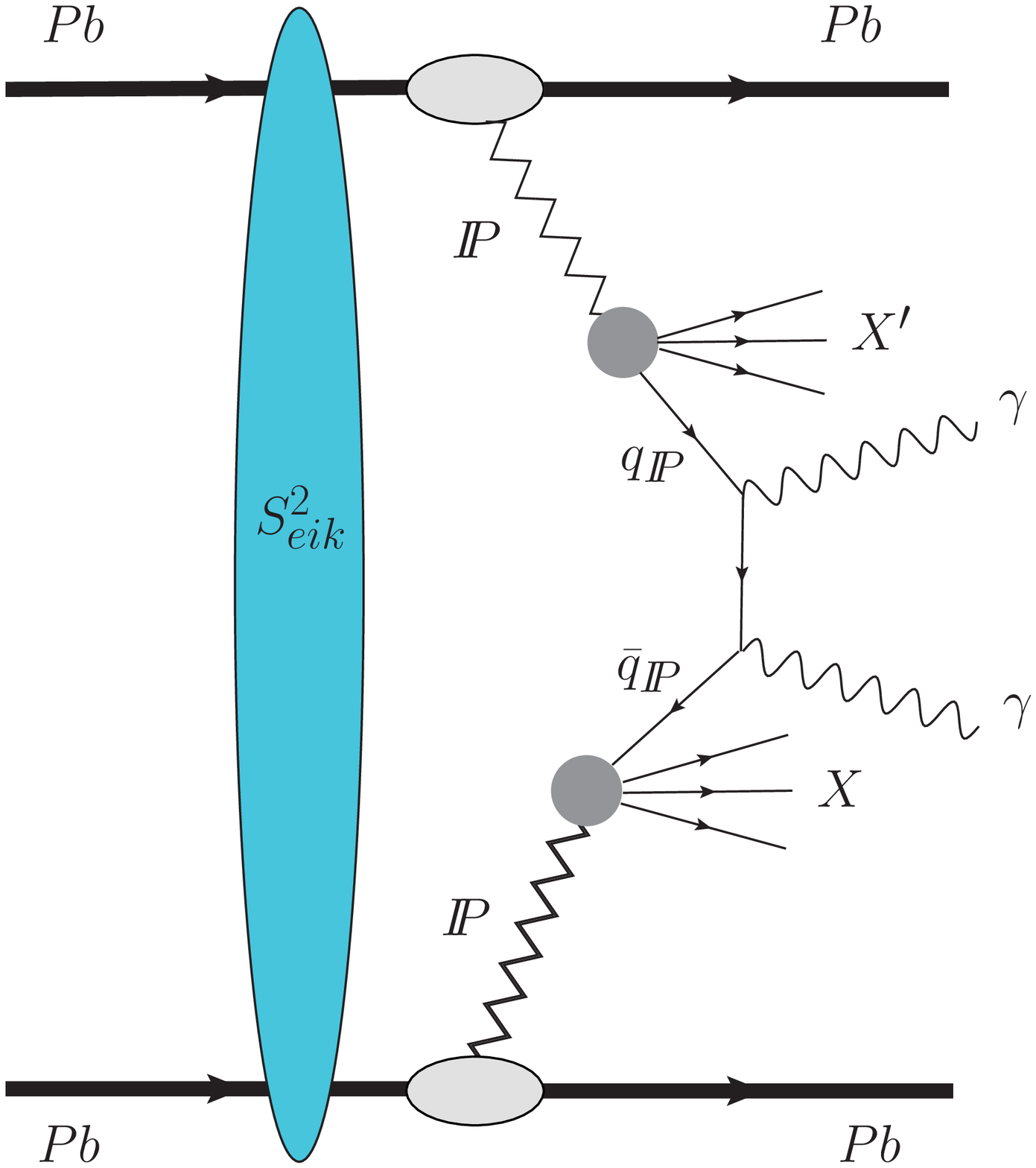,width=4.2cm}} \\
(c) & (d) \\ 
%{\psfig{figure=pp-psi-psi-500.eps,width=5cm}} & 
%{\psfig{figure=pp-psi-psi-7000.eps,width=5cm}} & 
%{\psfig{figure=pp-psi-psi-14000.eps,width=5cm}}
\end{tabular}                                                                                                                       
\caption{Diphoton production in $PbPb$ collisions by (a) the Light -- by -- Light scattering, (b) the central exclusive process induced by gluons (Durham process), and the double diffractive processes induced by (c) gluons and (d) quarks of the Pomeron ($\pom$).}
\label{fig:diagram}
\end{figure}

Light-by-light (LbL) scattering is a very rare phenomenon in which two photons interact, producing another pair of photons. This process was one of the most important predictions in the beginning of Quantum Electrodynamics (QED), and has no parallel in classical electrodynamics theory. 
Direct evidence for light-by-light scattering at high energy had proven difficulty to detect for decades. Although LbL scattering via an electron loop has been indirectly tested through the high precision measurements of the electron and muon anomalous magnetic moment \cite{VanDyck101103,g2collab101103}, direct observations in the laboratory remained inconclusive until recently, when the CMS and ATLAS Collaboration have observed, for the first time, the light -- by -- light (LbL) scattering  in ultraperipheral PbPb Collisions   \cite{Aad:2019ock,Sirunyan:2018fhl}.
Such collisions are characterized by an impact parameter $b$ greater than the sum of the radius of the colliding  nuclei \cite{upc1,upc2,upc3,upc4,upc5,upc6,upc7,upc8,upc9} and by a   photon -- photon luminosity that scales with $Z^4$, where $Z$ is number of protons in the nucleus. As a consequence, in ultraperipheral heavy ion collisions (UPHIC),   the elementary  elastic  $\gamma \gamma \rightarrow \gamma \gamma$ process, which  occurs at one -- loop level at order $\alpha^4$ and have a tiny cross section, is enhanced by a large $Z^4$ ($\approx 45 \times 10^6$) factor, becoming it feasible for the experimental analysis \cite{gustavo,antoni}. The LbL scattering in ultraperipheral $PbPb$ collisions is represented in Fig. \ref{fig:diagram} (a), with the resulting  final state being very clean, consisting  of the diphoton system,  two intact nuclei and  two rapidity gaps, i.e. empty regions  in pseudo-rapidity that separate the intact very forward nuclei from the $\gamma \gamma$ system. The recent experimental results have  motivated a series of studies that propose the analysis of the diphoton production in  $\gamma \gamma$ interactions as a probe of Beyond Standard Model (BSM) physics (See e.g. Refs.\cite{knapen,royon,gomes,liu}). However, in order to be possible to search by New Physics in $\gamma \gamma$ channel, it is fundamental to have control of background associated to other production channels that also generated a similar final state. Two potential backgrounds are the diphoton production in  central exclusive processes induced by gluons, represented in Fig. \ref{fig:diagram} (b) and denoted Durham process hereafter, and in double diffractive processes, represented in Figs. \ref{fig:diagram} (c) and (d). Such reactions also are characterized by two rapidity gaps and two intact ions in the final state,  but the diphoton system  is generated by the interaction between gluons  of the nucleus (Durham process) or gluons of the Pomeron ($\pom$), which is a color singlet object inside the nucleus, in the case of double diffractive processes.
One of goals of this paper is to estimate the contribution of each one of these production channels taking into account the acceptance of the LHC detectors. In particular, we will consider the typical set of cuts used by the ATLAS and CMS Collaborations to separate the exclusive events. In addition, we will present, for the first time, a detailed comparison between these distinct channels for the diphoton production in the kinematical range probed by the LHCb detector (For a previous study of the LbL scattering at the LHCb see Ref. \cite{mariola_ronan}). We will explore the possibility present in this detector  of probe diphotons with small invariant mass. A second goal of our paper is to present, for the first time, predictions for the diphoton production in $PbPb$ collisions for the energies of the   High -- Energy LHC ($\sqrt{s} = 10.6$ TeV) \cite{he_lhc} and Future Circular Collider ($\sqrt{s} = 39$ TeV) \cite{fcc}. In order to obtain the results for these future colliders we will consider the typical configurations of  central and forward detectors and similar cuts to those used to LHC. As we will demonstrate below, the possibility of probe the LbL scattering in the LHCb detector is very promising, as well at the HE -- LHC and FCC. Our results indicate that the background associated to the Durham and double diffractive processes can be strongly suppressed, which will allow  to perform a detailed  study of the LbL scattering as well the search of BSM physics using this final state.

This paper is organized as follows. In the next Section, we present a brief review of the formalism used to describe the  diphoton production in $PbPb$ collisions by the LbL, Durham and double diffractive processes. Moreover, we discuss the treatment of the soft survival effects.  In Section \ref{sec:res}, we present our results for the $\gamma \gamma$ production at the LHC, HE -- LHC and FCC. Predictions for cross sections and  the invariant mass, rapidity, transverse momentum  and acoplanarity distributions are presented. The impact of the selection cuts is discussed and predictions for a typical central and forward detector are presented.  Finally, in Section \ref{sec:conc}, our main conclusions are summarized.

\section{Formalism}

Initially, we  present a brief review of the main formulas to describe the exclusive diphoton production by $\gamma \gamma$ interactions in ultraperipheral $PbPb$ collisions, represented in Fig. \ref{fig:diagram} (a).  
 Assuming the impact parameter representation and considering  the Equivalent Photon Approximation (EPA) \cite{epa}, the total cross section can be factorized in terms of the equivalent photon spectrum  
of the incident nuclei and the elementary cross section for the elastic $\gamma \gamma \rightarrow \gamma \gamma$ process as follows
\begin{eqnarray}
\sigma \left(Pb Pb \rightarrow Pb \otimes \gamma \gamma \otimes Pb;s \right)   
&=& \int \mbox{d}^{2} {\mathbf r_{1}}
\mbox{d}^{2} {\mathbf r_{2}} 
\mbox{d}W 
\mbox{d}Y \frac{W}{2} \, \hat{\sigma}\left(\gamma \gamma \rightarrow \gamma \gamma ; 
W \right )  N\left(\omega_{1},{\mathbf r_{1}}  \right )
 N\left(\omega_{2},{\mathbf r_{2}}  \right ) S^2_{abs}({\mathbf b})  
  \,\,\, .
\label{cross-sec-2}
\end{eqnarray}
where $\sqrt{s}$ is center - of - mass energy of the $PbPb$ collision, $\otimes$ characterizes a rapidity gap in the final state, 
$W = \sqrt{4 \omega_1 \omega_2} = m_X$ is the invariant mass of the $\gamma \gamma$ system and $Y = y_{\gamma \gamma}$ its rapidity. The photon energies $\omega_1$ and $\omega_2$  are related to   
$W$ and to the rapidity  $Y$ of the outgoing diphoton system by 
\begin{eqnarray}
\omega_1 = \frac{W}{2} e^Y \,\,\,\,\mbox{and}\,\,\,\,\omega_2 = \frac{W}{2} e^{-Y} \,\,\,.
\label{ome}
\end{eqnarray}
The cross section $\hat{\sigma}$ is the elementary cross section to produce a pair of photons, which will be calculated taking into account of the fermion loop contributions as well as  the contribution from $W$ bosons. Moreover, $N(\omega_i, {\mathbf r}_i)$ is the equivalent photon spectrum   with energy $\omega_i$ at a transverse distance ${\mathbf r}_i$  from the center of nucleus, defined in the plane transverse to the trajectory, which is determined by the charge form factor of the nucleus. Finally,  the factor $S^2_{abs}({\mathbf b})$ depends on the impact parameter ${\mathbf b}$ of the $PbPb$ collision and  is denoted the absorptive  factor, which excludes the overlap between the colliding nuclei and allows to take into account only ultraperipheral collisions. Currently, there are different approaches to treat these soft survival corrections. For example, 
Baur and Ferreira - Filho \cite{Baur_Ferreira} have proposed to exclude the 
strong interactions between the incident nuclei by assuming that 
\begin{eqnarray}
S^2_{abs}({\mathbf b}) = \Theta\left(
\left|{\mathbf b}\right| - 2 R
 \right )  = 
\Theta\left(
\left|{\mathbf b_{1}} - {\mathbf b_{2}}  \right| - 2 R
 \right )  \,\,,
\label{abs1}
\end{eqnarray}
where $R$ is the nuclear radius. Such equation treats the nuclei as hard spheres with radius $R$ and assumes that the probability to have a hadronic interaction when $b > 2 R$ is zero.
On the other hand,  in the 
STARLight  \cite{starlight} and SuperChic \cite{superchic3} event generators, the authors have proposed  distinct models based on the Glauber formalism. We have verified that for small values of $W$, which is the focus of the analysis performed in this paper, the predictions of these different approaches are almost identical, in agreement with the analysis performed in  Ref. \cite{celsina} where  the authors have presented a detailed analysis about the description of exclusive $\gamma \gamma$ interactions in $PbPb$ collisions considering different models for the form factor  and for the absorptive corrections. 

For the exclusive production of a diphoton in the gluon -- induced  interactions represented in Fig. \ref{fig:diagram} (b),  we will consider the model proposed by Khoze, Martin and Ryskin~\cite{Khoze:2000cy, Khoze:2001xm, Khoze:2000jm} some years ago, denoted Durham model hereafter, which has been used to estimate a large number of different final states and have  predictions in reasonable agreement with the observed rates for exclusive processes measured by the CDF collaboration~\cite{Aaltonen:2007am,Aaltonen:2007hs,Aaltonen:2009kg} and in the Run I of the LHC (For a recent review see Ref. \cite{review_lang}). In this model, the amplitude for the diphoton production in a $pp$ collision can be  expressed  as follows
\begin{equation}
 {\cal{A}}_{pp}(s,q_{1\perp},q_{2\perp}) = \pi^2 \int \frac{d^2{Q}_{\perp} \, \bar{{\cal{M}}}}{{Q}_{\perp}^2   q_{1\perp}^2 q_{2 \perp}^2} f_g(x_1,x_1^{\prime},Q_{\perp}^2, \mu^2;t_1) f_g(x_2,x_2^{\prime},Q_{\perp}^2, \mu^2;t_2)
\label{eq:kmr}
\end{equation}
where  $Q_{\perp}^2$ is the virtuality of the soft gluon needed for color  screening, $q_{1 \perp}$ and $q_{2 \perp}$ are the  transverse momenta of the gluons which participate of the hard subprocess  and
$x_i$ ($x_i^{\prime}$) are the momentum fractions carried by the fusing (screening) gluons. Moreover, 
$\bar{{\cal{M}}}$ is the color -- averaged, normalized sub -- amplitude for the $gg \rightarrow \gamma \gamma$ subprocess. The quantities $f_g$ are the  skewed unintegrated gluon densities evaluated to the factorization scale $\mu$, which we assume to be of the order of the invariant mass $m_X$ of the final state. The $t$ -- dependence of the skewed distribution is assumed to factorize out as a proton form factor, being $\propto \exp(b t/2)$ with $b = 4$ GeV$^{-2}$. Since 
\begin{eqnarray}
(x^{\prime} \approx \frac{Q_{\perp}}{\sqrt{s}}) \ll (x \approx \frac{m_{X}}{\sqrt{s}}) \ll 1
\end{eqnarray}
 it is possible to express $f_g(x,x^{\prime},Q_{\perp}^2, \mu^2)$, to single log accuracy, in terms of the conventional integrated gluon density $g(x)$, together with a known Sudakov suppression $T$ which ensures that the active gluons do not radiate in the evolution from $Q_{\perp}$ up to the hard scale $\mu \approx m_{X}/2$. In this paper we will calculate  $f_g$ in the proton case considering that  the integrated gluon distribution $xg$ is described by the MMHT parametrization \cite{mmht}.
The Eq. (\ref{eq:kmr}) corresponds to the amplitude for the exclusive production of a diphoton in a hard process without no further perturbative emission. However, the exclusivity of the event can be spoiled by secondary particles that can be produced by additional soft hadronic interactions.  
 Such  soft survival effects are, in general, parametrized in terms of a rapidity gap survival probability, $S^2$, which corresponds to the probability of the scattered proton not to dissociate due to the secondary interactions.  In Ref. \cite{superchic3} the Durham model was generalized  for ion  -- ion collisions by proposing to express the scattering amplitude for the coherent $A_1 A_2$ process in terms of the $pp$ amplitude described above and the nuclear form factors $F_{A_i}$ as follows
\be\label{eq:tqcdex}
{\cal{A}}_{A_1 A_2}(s,q_{1\perp},q_{2\perp}) = {\cal{A}}_{pp}(s,q_{1\perp},q_{2\perp}) F_{A_1}(Q_1^2) F_{A_2}(Q_2^2)\;,
\ee
where  $Q^2_i={(q_{i_\perp}^2+x_i^2 m_{N_i}^2)}/({1-x_i})$.
Such equation was written in the impact parameter space and a model for the soft survival effects was included in the calculation. The resulting ion -- ion cross section is proportional to the $pp$ one and to the nuclear opacity, which encodes the probability for no additional ion -- ion rescattering at different impact parameters.  One important aspect is that the nuclear version of the Durham model is implemented in the  SuperChic3 Monte Carlo event generator, being possible to perform the analysis with and without the inclusion of the soft survival effects.  

%The Durham group have proposed an approach to treat the soft survival effects associated to the additional soft proton -- proton interactions (denoted eikonal factor $S^2_{{eik}}$), which are independent of the hard processes, as well the rescatterings of the protons with the intermediate partons that are described by the so--called enhanced factor $S^2_{{enh}}$. As discussed in Ref. \cite{khoze}, the magnitude of the enhanced factor is still uncertain, but it is expected to generate a weaker suppression in comparison to that associated to the eikonal survival factor. Consequently, such corrections will be not included in our analysis.

Finally, for the description of the diphoton production in the double diffractive processes (DDP) represented in Figs. \ref{fig:diagram} (c) and (d),  we will consider the Resolved Pomeron model, in which the {pomeron} is assumed to have a partonic structure \cite{IS}. We have that 
the corresponding cross section can be expressed by
\begin{eqnarray}
\sigma(Pb Pb \rightarrow Pb \otimes X +  \gamma \gamma +  X^{\prime} \otimes Pb) & = &  \left\{ \int dx_{1} \int dx_{2} \, \left[g^D_{1}(x_{1},\mu^2) \cdot g^D_{2}(x_{2},\mu^2) \cdot \hat{\sigma}(g g \rightarrow \gamma \gamma)\right.\right.  \nonumber \\
& + & \left. \left. \, [q^D_{1}(x_{1},\mu^2) \cdot \bar{q}^D_{2}(x_{2},\mu^2) + \bar{q}^D_{1}(x_{1},\mu^2) \cdot {q}^D_{2}(x_{2},\mu^2)]  \cdot \hat{\sigma}(q \bar{q} \rightarrow \gamma \gamma)\right]\right\} \,\,,
\label{pompom}
\end{eqnarray}
where $g^D_i (x_i,\mu^2)$, $q^D_i (x_i,\mu^2)$ and $\bar{q}^D_i (x_i,\mu^2)$ are the diffractive gluon, quark and antiquark densities of the nucleus $i$ with a momentum fraction $x_i$. In the  Resolved Pomeron model \cite{IS} the diffractive parton distributions  are expressed in terms of parton distributions in the {pomeron} and a Regge parametrization of the flux factor describing the {pomeron} emission by the hadron. The  parton distributions have its evolution given by the DGLAP evolution equations and should be determined from events with a rapidity gap or a intact hadron. 
In order to specify the diffractive distributions for a nucleus, we will follow the approach proposed in Ref. \cite{vadim} (See also Ref. \cite{review_vadim}). In this approach, the diffractive distributions for a nucleus are estimated  taking into account the nuclear effects associated to the nuclear coherence and the leading twist nuclear shadowing. The basic assumption is that  the  pomeron - nucleus coupling is proportional to the mass number {A} \cite{berndt}. As the associated pomeron flux depends on the square of this coupling, this model predicts that the pomerons are coherently emitted by the nucleus, which implies that the {pomeron} flux emitted by the nucleus,  $f_{\pom/{\rm A}}$, is proportional to ${\rm A}^2$.  Consequently, the nuclear diffractive gluon distribution can be expressed as follows (For details see Ref. \cite{vadim})
\begin{eqnarray}
{ g^D_{\rm A}(x,\mu^2)}= R_g \, {\rm A}^2 \,  { \int_x^1 \frac{dx_{\pom}}{x_{\pom}} \left[ \int dt \, f_{\pom/{\rm p}}(x_{{\pom}}, t) \cdot F_{\rm A}^2(t) \right]  g_{\pom}\left(\frac{x}{x_{\pom}}, \mu^2\right)} \,\,,
\label{difgluon:nucleo}
\end{eqnarray}
where $R_g$ is the suppression factor associated to the nuclear shadowing, 
$f_{\pom/\rm p}(x_{\pom},t)$ is the {pomeron} flux emitted by the proton and $g_{\pom}(\beta, \mu^2)$ is the gluon distribution in the {pomeron}, with  $\beta$  being the momentum fraction carried by the partons inside the {pomeron}.  Moreover,  $F_{\rm A}(t)$ is the nuclear form factor. A similar relation is also valid for the diffractive quark and antiquark densities of the nucleus. In what follows we will assume that $R_g  = 0.15$ as in Ref. \cite{vadim} and that $F_{\rm A}(t) \propto e^{R_{\rm A}^2 t/6}$, with $R_{\rm A}$ being the nuclear radius. It is important to emphasize that our group have implemented this generalization in the Forward Physics Monte Carlo (FPMC) \cite{fpmc}, which allow us to estimate the associated cross sections and distributions taking into account of the detector acceptances.

Similarly to the exclusive case, the predictions for the diphoton production in double diffractive processes are also expected to be strongly modified  by 
soft interactions which lead to an extra production of particles that destroy the rapidity gaps in the final state \cite{bjorken}. As these effects have nonperturbative nature, they are difficult to treat and its magnitude is strongly model dependent (For recent reviews see Refs. \cite{durham,telaviv}).
In our analysis of the soft survival corrections in ion -- ion collisions we will assume that them can be factorized of the hard process and that its effects can be included in the calculation by multiplying the cross section by  a global factor $S^2_{eik}$ (denoted eikonal factor). In order to estimate $S^2_{eik}$, we will   consider the approach proposed in Ref. \cite{nos_dijet}, which generalizes the model described in Ref. \cite{berndt} for coherent double exchange processes in nuclear collisions. The basic idea in this approach is to express the double diffractive cross section in the impact parameter space, which implies that it  becomes dependent on the magnitude of the geometrical overlap of the two nuclei during the collision. As a consequence, it is possible to take into account the centrality of the incident particles and estimate the absorptive corrections associated to the additional soft hadronic interactions  by requiring that the colliding nuclei remain intact, which is equivalent to suppress  the interactions at small impact parameters ($b < R_{\rm A} + R_{\rm B}$). A detailed description of this approach is presented in Appendix A of Ref. \cite{nos_dijet}. The resulting predictions for $PbPb$ collisions at $\sqrt{s} = 5.5, \, 10.6$ and 39 TeV are $3.4 \times 10^{-5}$, $2.1 \times 10^{-5}$ and $1.0 \times 10^{-5}$, respectively.
It is important to emphasize that these predictions are larger than those obtained in Ref. \cite{miller} using a Glauber approach and in Ref. \cite{radion} assuming that the nuclear suppression factor is given by $ {\mathcal S}^2_{{\rm A}_1{\rm A}_2} = {\mathcal S}^2_{\rm pp}/({\rm A}_1.{\rm A}_2)$.  We  will consider that this approach can also be used to include the soft survival effects in the exclusive diphoton process represented by the  Fig. \ref{fig:diagram} (b). In other words, we will estimate the exclusive diphoton using the SuperChic MC and the double diffractive process using the FPMC and  its predictions will be multiplied by the same factor $S^2_{{eik}}$, which is energy dependent.
We have verified that our predictions for the Durham model are one order of magnitude larger than those obtained considering the survival model implemented in the SuperChic MC. Consequently, our predictions for the diphoton production in the central exclusive [Fig. \ref{fig:diagram} (b)]  and  the double diffractive [Figs. \ref{fig:diagram} (c) and (d)]  processes may be considered an upper bound for the magnitude of these cross sections.

\begin{center}
\begin{table}[t!]
\begin{tabularx}{\textwidth}{@{}l *3{>{\centering\arraybackslash}X}@{}}
\hline\hline
\multirow{2}{*}{Process} & \multirow{2}{*}{$\sqrt{s}$ (TeV)} & \multirow{2}{*}{$\sigma[Pb\:Pb\to  Pb+ \gamma \gamma +Pb ]$} \\
         &               &             \\
            \hline
\multirow{2}{*}{LbL}
        & 5.5                       &$1.8\times10^{4}$~nb    \\
        & 10.6                       &$2.7\times10^{4}$~nb    \\ 
        & 39                       & $5.2\times10^{4}$~nb  \\ 
          \hline
\multirow{2}{*}{Durham}
        & 5.5                       &$4.9\times10^{6}$~nb (167.0 ~nb)                  \\
          & 10.6                    &$9.8\times10^{6}$~nb  (333.2 ~nb)    \\ 
        & 39                       &  $3.8\times10^{7}$~nb  (380.0 ~nb) &    \\ 
\hline
\multirow{2}{*}{DDP}
        & 5.5                        &  $5.2\times10^{5}$~nb (17.7 ~nb)       \\
  & 10.6                       &$9.7\times10^{5}$~nb   (22.3 ~nb) \\ 
        & 39                       & $3.0\times10^{6}$~nb (30.0 ~nb)          \\
\hline\hline
\end{tabularx}
\caption{\label{tab:generation} Predictions for the diphoton production cross sections in $PbPb$ collisions at LHC, HE -- LHC and FCC. The results in the parenthesis are the predictions after the inclusion of soft survival factor $S^2_{{eik}}$.}
\end{table}
\end{center}

 \begin{center}
 \begin{figure}[htbp!]
 \includegraphics[width=0.45\textwidth]{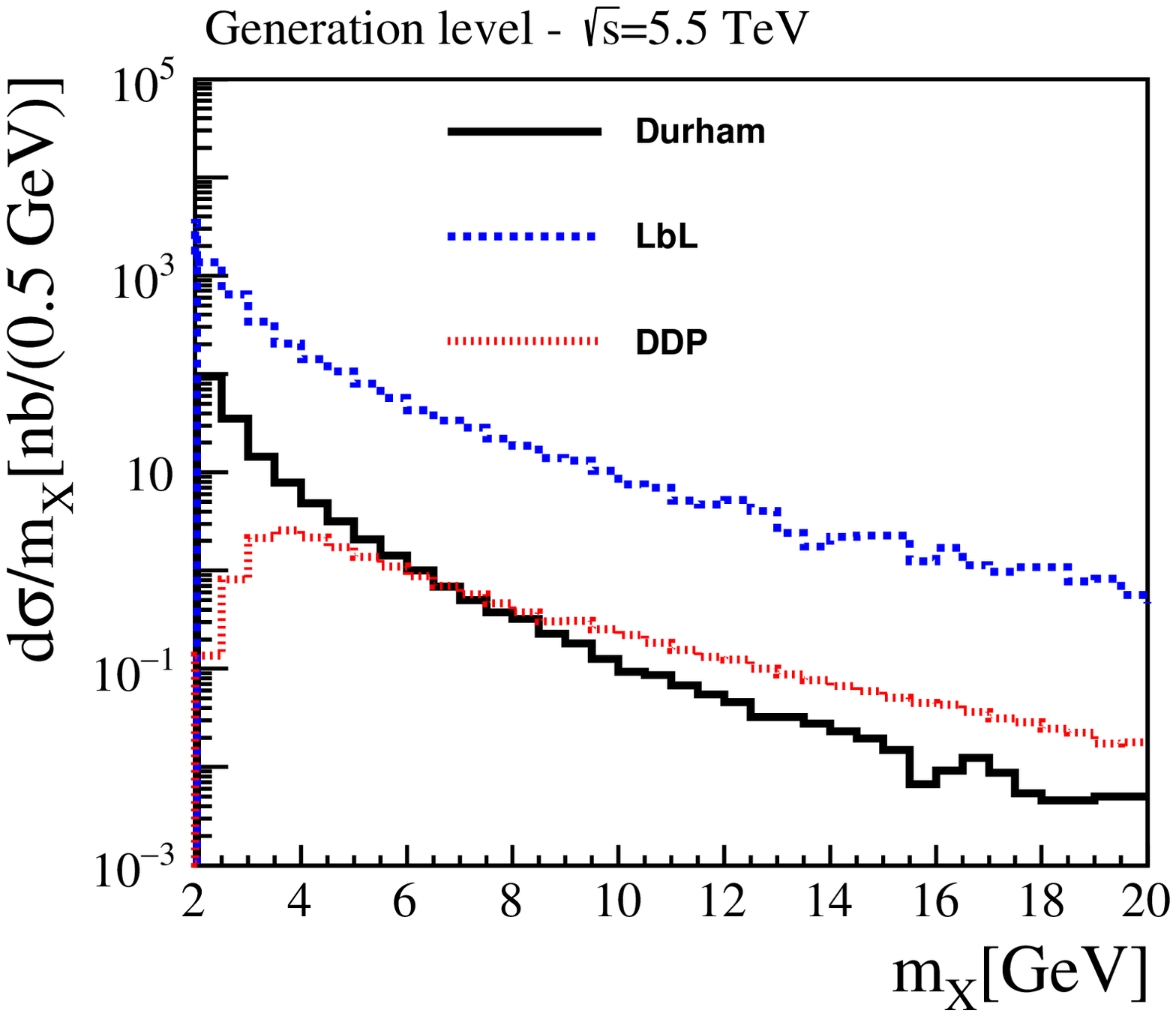}
 \includegraphics[width=0.45\textwidth]{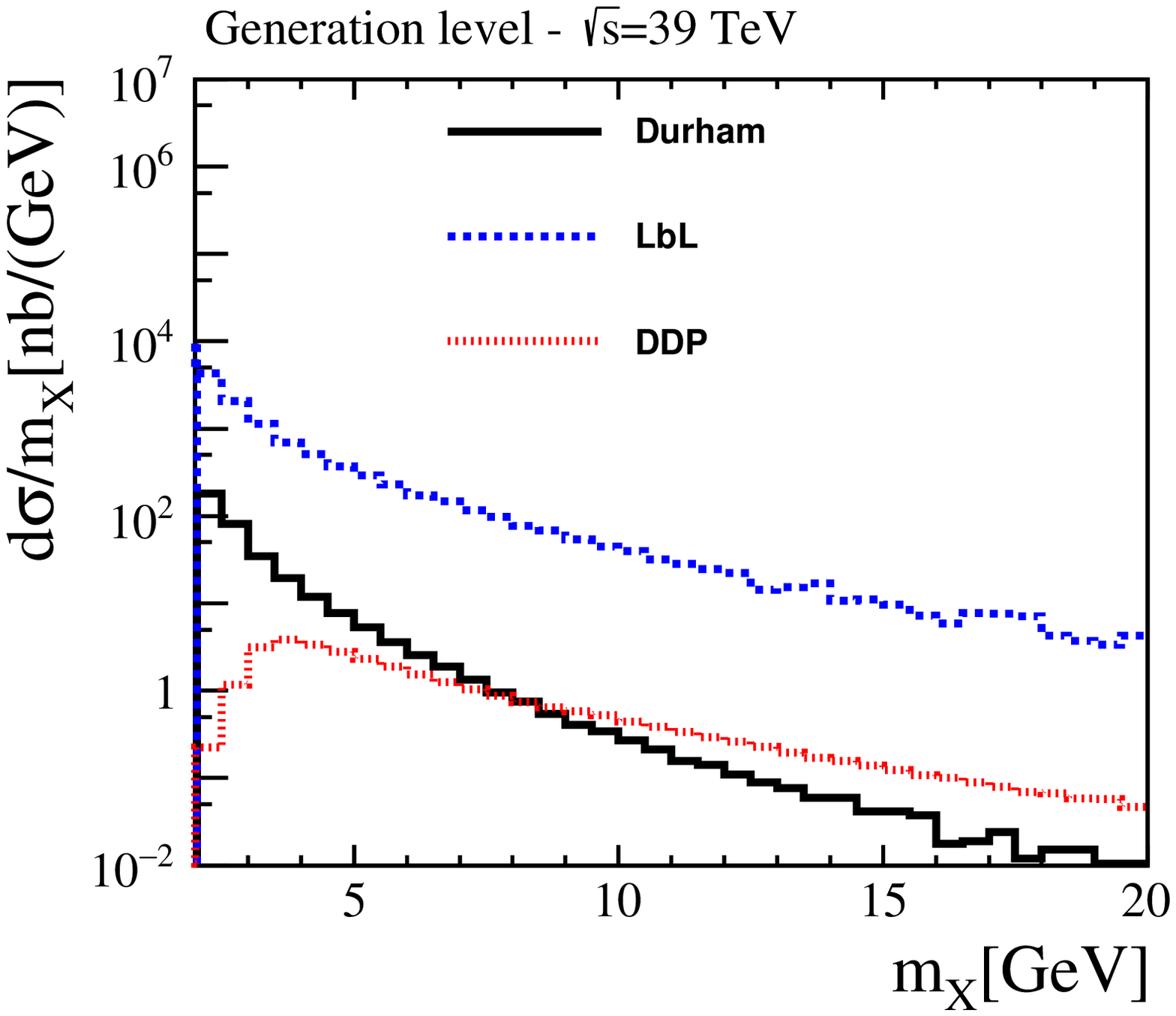}
 \includegraphics[width=0.45\textwidth]{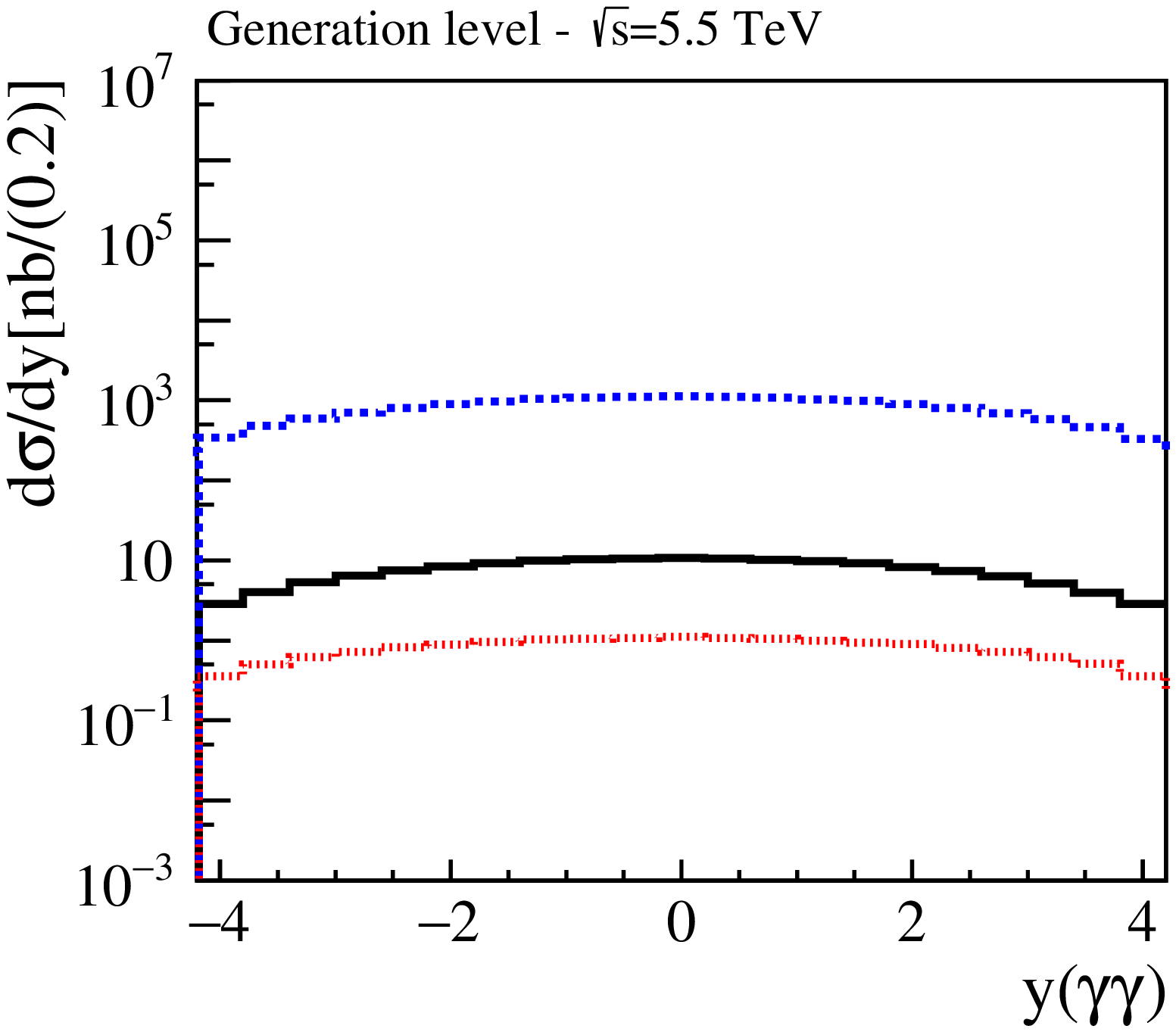}
 \includegraphics[width=0.45\textwidth]{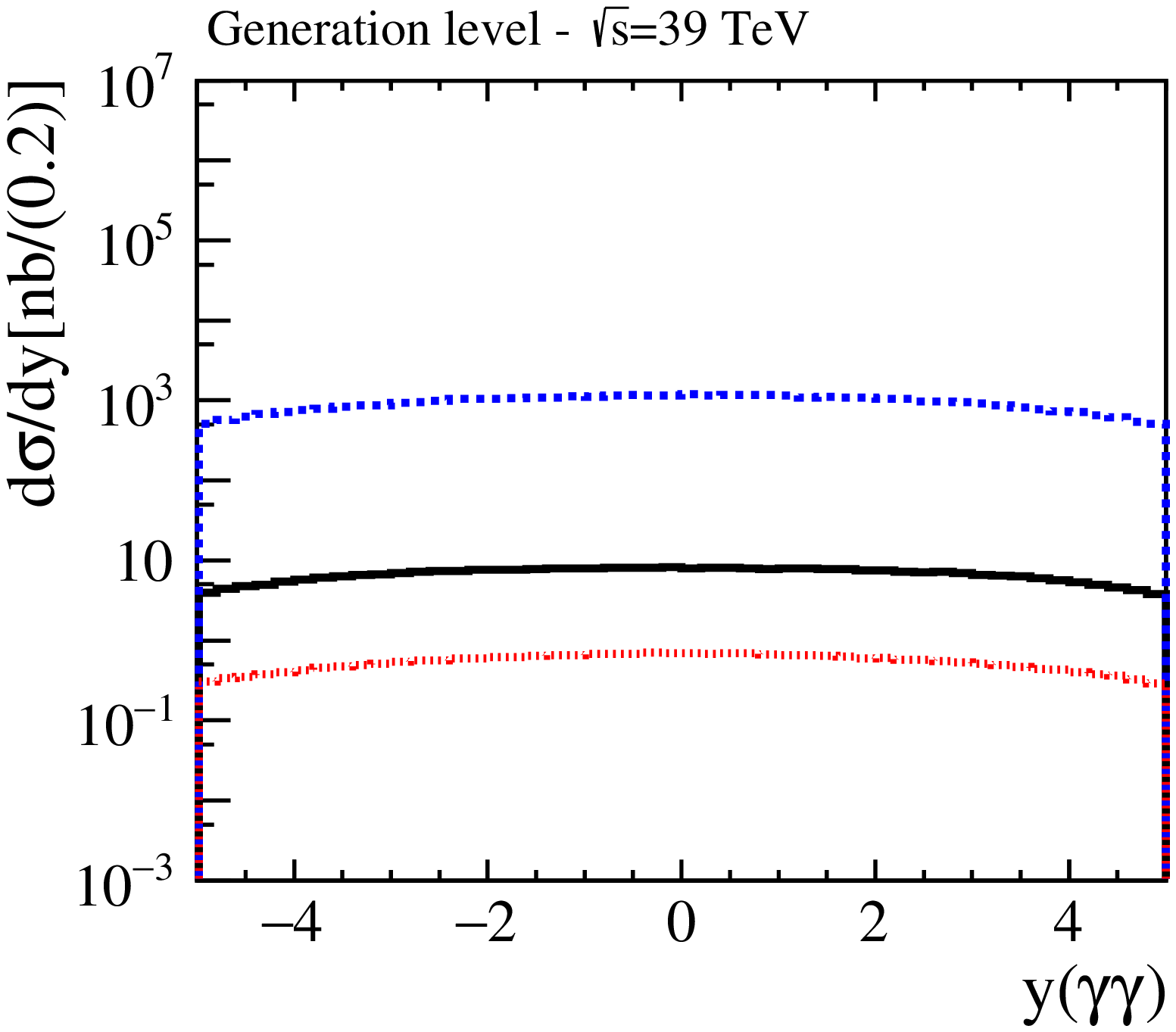}
\caption{Predictions for the invariant mass $m_X$ and rapidity $y_{\gamma \gamma}$ distributions of the diphoton system produced in $PbPb$ collisions at the LHC (left panels) and FCC (right panels). Results at the generation level, without the inclusion of experimental cuts.}
 \label{fig:generation}
 \end{figure}
 \end{center}

\section{Results}
\label{sec:res}

In what follows we will present our results for the exclusive and diffractive diphoton production in $PbPb$ collisions at $\sqrt{s} = 5.5$, 10.6 and 39 TeV. In our analysis we will use the SuperChic MC event generator \cite{superchic3} to estimate the processes represented in the Figs. \ref{fig:diagram} (a) and (b). On the other hand, the double diffractive diphoton production [Figs. \ref{fig:diagram} (c) and (d)], will be calculated considering the FPMC event generator \cite{fpmc}. Initially, in Table \ref{tab:generation} we  present our results for the  cross sections associated to the different channels, obtained at the generation level, without the inclusion of any selection in the events. We have that the gluon -- induced processes (Durham and DDP) are strongly suppressed by the soft survival effects, with the associated cross sections being a factor $\ge 10^2$ smaller than the photon -- induced one (LbL). Moreover, the DDP predictions are one order of magnitude smaller than the Durham one. In Fig. 
\ref{fig:generation} we present our results for the invariant mass ($m_X$) and the rapidity ($y_{\gamma \gamma}$) distributions of the diphoton system for $PbPb$ collisions at the LHC (left panels) and FCC (right panels). One have that the LbL dominates the diphoton production in the $m_X$ and $y_{\gamma \gamma}$ ranges considered. In addition, our results indicate that the
double diffractive prediction is larger than the Durham one for large values of the invariant mass, but is strongly suppressed at small -- $m_X$. 

In order to obtain realistic estimates of the diphoton production in $PbPb$ collisions, which can be compared with the future experimental data, we will include in our analysis the experimental cuts  that are expected to be feasible in the next run of the LHC and in the future at the HE -- LHC and FCC. We will consider two distinct configurations of cuts: one for a typical central detector as ATLAS and CMS, and other for a forward detector as  LHCb. In particular, we will analyze the possibility of study diphotons with invariant mass in the range $1 \le m_X \le 5$ GeV using the LHCb detector. Currently, such low mass range  cannot be reached by the central detectors. The selection criteria implemented in our analysis of double diffractive and exclusive diphoton processes are the following:   
\begin{itemize}
\item For a central detector: We will select events in which $m_{X}$ > 5 GeV and $E_{T}(\gamma,\gamma)$ > 2 GeV, where $E_T$ is the transverse energy of the photons. Moreover, we will impose a cut on  the acoplanarity ($1 -(\Delta \phi/\pi)$ < 0.01) and  transverse momentum of the diphoton system  ($p_{T}(\gamma,\gamma)$ < 0.1 GeV). Finally, we only will select events where photons are produced in the rapidity range $|\eta(\gamma^{1},\gamma^{2})| < 2.5$ with 0 extra tracks.

\item For a forward detector: We will select events in which $m_{X}$ > 1 GeV and $p_{T}(\gamma,\gamma)$ > 0.2 GeV, where $p_T$ is the transverse momentum of the photons. Moreover, we will impose a cut on  the acoplanarity ($1 -(\Delta \phi/\pi)$ < 0.01) and  transverse momentum of the diphoton system  ($p_{T}(\gamma,\gamma)$ < 0.1 GeV). Finally, we only will select events where photons are produced in the rapidity range $2.0 < |\eta(\gamma^{1},\gamma^{2})| < 4.5$ with 0 extra tracks with $p_{T} > 0.1$ GeV in the rapidity range $-3.5 < \eta < -1.5$ and
$p_T > 0.5$ GeV in the range $-8.0 < \eta < -5.5$.
\end{itemize}

\begin{center}
\begin{table}
\begin{tabular}{|c|c|c|c|}
\hline 
{\bf $PbPb$ collisions at $\sqrt{s}$ = 5.5 TeV} & LbL & Durham & DDP \tabularnewline
\hline 
Total Cross section {[}nb{]} &18000.0  & 167.0  & 17.7  \tabularnewline
\hline 
$m_{X}> 5\:\rm{GeV}, E_{T}(\gamma,\gamma)>2\:\rm{GeV}$& 187.0  & 3.6   &17.7    \tabularnewline
\hline 
$1- (\Delta \phi/\pi) < 0.01$ & 186.0  & 3.1 &6.9   \tabularnewline
\hline 
$p_{T}(\gamma\gamma)< 0.1$  GeV & 139.0 & 2.8 & 0.1  \tabularnewline
\hline 
 $|\eta(\gamma,\gamma)|<2.5$ & 139.0 & 1.9 &  0.0  \tabularnewline
\hline
%\hline 
%$13 < m\left(\gamma\gamma\right) < 17 $  & 7.3  & 0.1 &0  &8.6  &-  \tabularnewline
%\hline 
%$38 < m\left(\gamma\gamma\right) < 42 $  & 0.5 & 0 &0  &- & 11.5   \tabularnewline
%\hline 
\hline 
\hline
{\bf $PbPb$ collisions at $\sqrt{s}$ = 10.6 TeV} &  LbL & Durham & DDP   \tabularnewline
\hline 
Total Cross section {[}nb{]} &27000.0  & 333.2  & 22.3  \tabularnewline
\hline 
$m_{X}> 5\:\rm{GeV}, E_{T}(\gamma,\gamma)>2\:\rm{GeV}$& 352.9  & 7.6   &13.5  \tabularnewline
\hline 
$1- (\Delta \phi/\pi) < 0.01$ & 352.8  & 6.7 & 0.1  \tabularnewline
\hline 
$p_{T}(\gamma\gamma)< 0.1$  GeV & 350.2 & 5.8 & 0.0  \tabularnewline
\hline 
 $|\eta(\gamma,\gamma)|<2.5$ & 227.6 &3.6  & 0.0  \tabularnewline
\hline
\hline 
{\bf $PbPb$ collisions  at $\sqrt{s}$ = 39 TeV} & LbL & Durham & DDP  \tabularnewline
\hline 
Total Cross section {[}nb{]} & 52000.0 &  380 & 30.0  \tabularnewline
\hline 
$m_{X}> 5\:\rm{GeV}, E_{T}(\gamma,\gamma)>2\:\rm{GeV}$& 844.0  & 9.2  &13.0  \tabularnewline
\hline 
$1- (\Delta \phi/\pi) < 0.01$ & 840.0 & 8.0 &0.1  \tabularnewline
\hline 
$p_{T}(\gamma\gamma)< 0.1$  GeV & 836.0 & 7.0 & 0.0 \tabularnewline
\hline 
 $|\eta(\gamma,\gamma)|<2.5$ & 431.0 & 3.4 &  0.0 \tabularnewline
\hline
\end{tabular}
\caption{Predictions for the central exclusive and double diffractive diphoton  cross sections after the inclusion of the exclusivity cuts for a typical central detector.}
\label{tab:central}
\end{table}
\end{center}

\begin{center}
\begin{table}
\begin{tabular}{|c|c|c|c|}
\hline 
{\bf $PbPb$ collisions at $\sqrt{s}$ = 5.5 TeV} & LbL & Durham & DDP\tabularnewline
\hline 
Total Cross section {[}nb{]} & 18000.0 & 167.0  & 17.7 \tabularnewline
\hline 
$m_{X}> 1\:\rm{GeV}, p_{T}(\gamma,\gamma)>0.2\:\rm{GeV}$& 13559.0 & 142.0 &17.6  \tabularnewline
\hline 
$1- (\Delta \phi/\pi) < 0.01$ & 8834.0 & 51.0 & 0.2 \tabularnewline
\hline 
$p_{T}(\gamma\gamma)< 0.1$  GeV & 8826.0 & 47.0 &0.0\tabularnewline
\hline 
 $2.0<\eta(\gamma,\gamma)<4.5$ &  616.0&  3.7& 0.0 \tabularnewline
\hline
\hline
{\bf $PbPb$ collisions at $\sqrt{s}$ = 10.6 TeV} & LbL & Durham & DDP \tabularnewline
\hline 
Total Cross section {[}nb{]} & 27000.0 & 333.2  &22.3  \tabularnewline
\hline 
$m_{X}> 1\:\rm{GeV}, p_{T}(\gamma,\gamma)>0.2\:\rm{GeV}$&20372.9  & 284.6  &22.0  \tabularnewline
\hline 
$1- (\Delta \phi/\pi) < 0.01$ &13958.5  & 103.2 & 0.3 \tabularnewline
\hline 
$p_{T}(\gamma\gamma)< 0.1$  GeV & 13949.0 & 95.1 & 0.0 \tabularnewline
\hline 
 $2.0<\eta(\gamma,\gamma)<4.5$ & 1069.5 & 8.3 & 0.0 \tabularnewline
\hline
\hline 
{\bf $PbPb$ collisions at $\sqrt{s}$ = 39 TeV} & LbL & Durham & DDP \tabularnewline
\hline 
Total Cross section {[}nb{]} & 52000.0 & 380.0   & 30.0   \tabularnewline
\hline 
$m_{X}> 1\:\rm{GeV}, p_{T}(\gamma,\gamma)>0.2\:\rm{GeV}$& 38025.0 & 325.0&30.0  \tabularnewline
\hline 
$1- (\Delta \phi/\pi) < 0.01$ &28216.0  &118.0  & 0.3 \tabularnewline
\hline 
$p_{T}(\gamma\gamma)< 0.1$  GeV & 28202.0 & 109.0 &0.0 \tabularnewline
\hline 
 $2.0<\eta(\gamma,\gamma)<4.5$ & 2229.0 &10.0  & 0.0 \tabularnewline
\hline

\end{tabular}
\caption{Predictions for the central exclusive and double diffractive diphoton  cross sections after the inclusion of the exclusivity cuts for a typical forward detector.}
\label{tab:forward}
\end{table}
\end{center}

 \begin{center}
 \begin{figure}[htbp!]
 \includegraphics[width=0.45\textwidth]{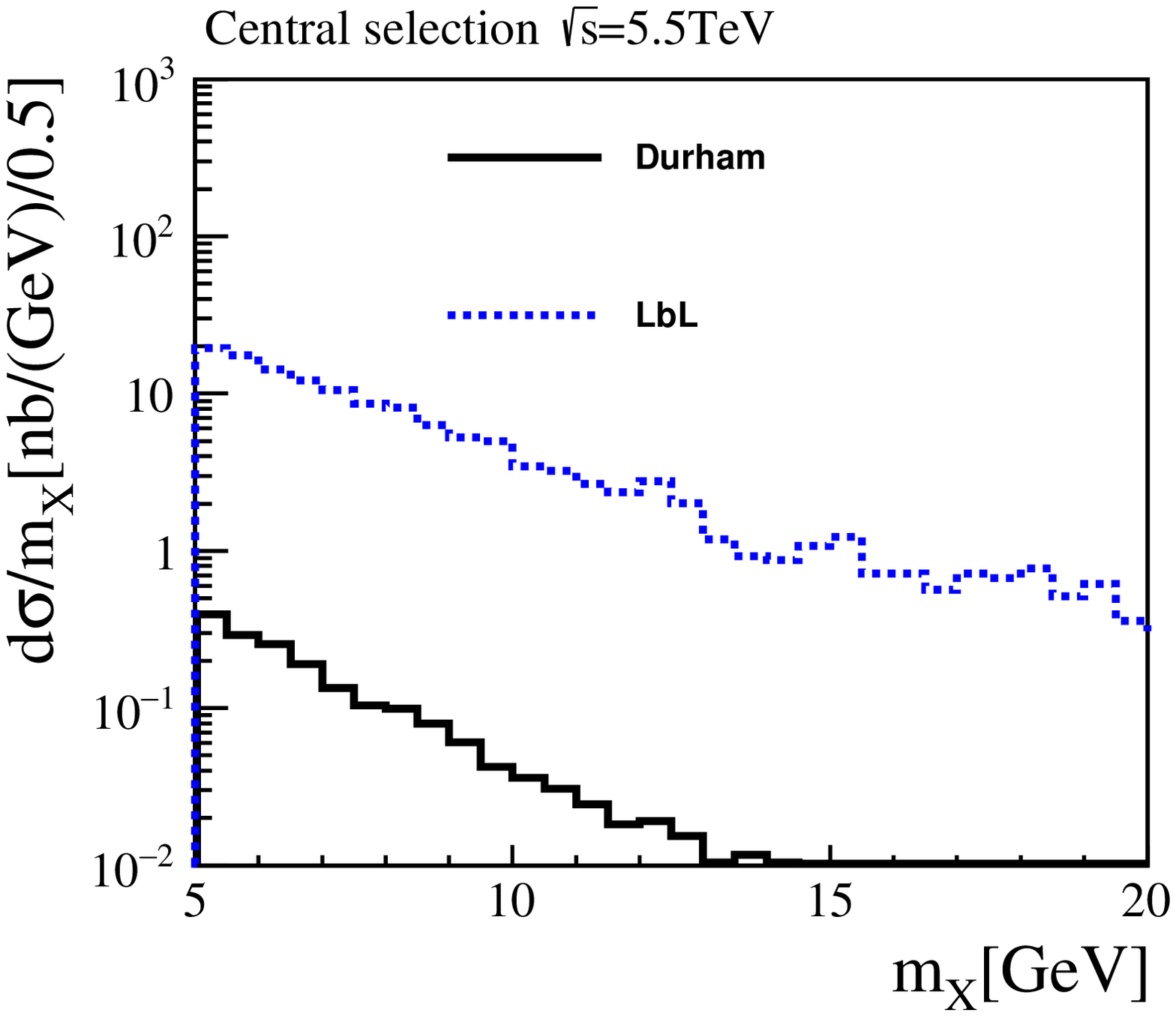}
 \includegraphics[width=0.45\textwidth]{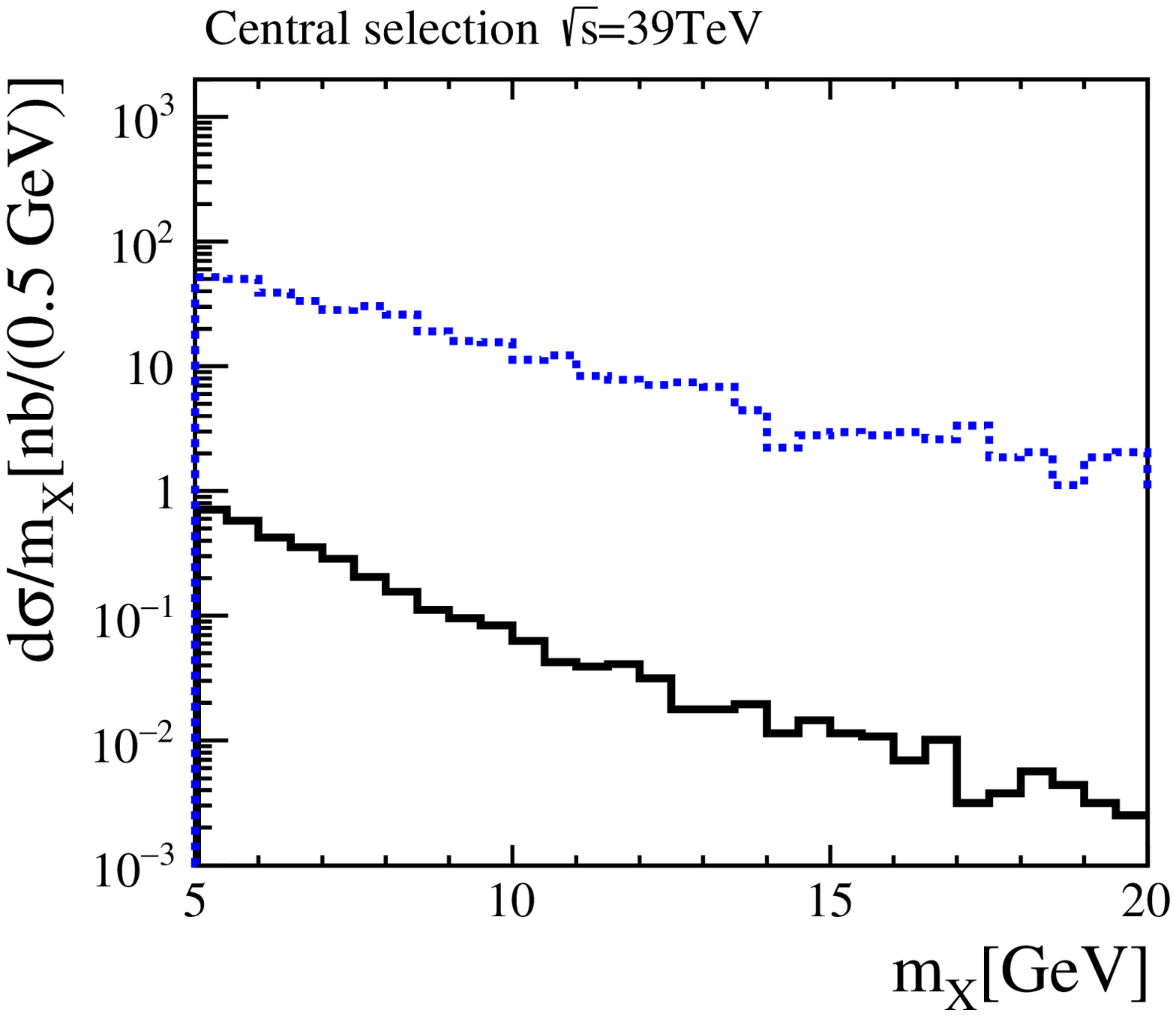}
 \includegraphics[width=0.45\textwidth]{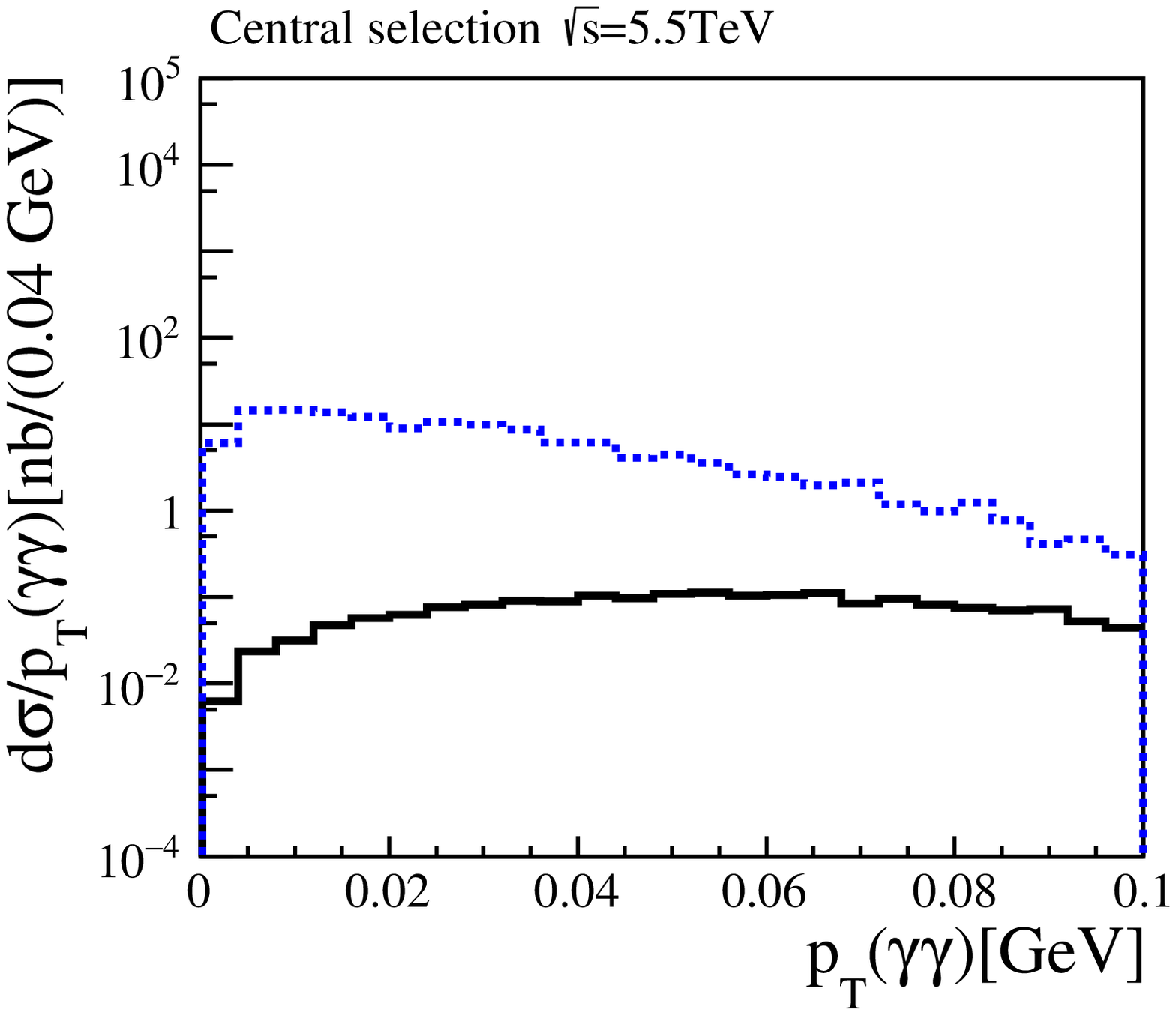}
 \includegraphics[width=0.45\textwidth]{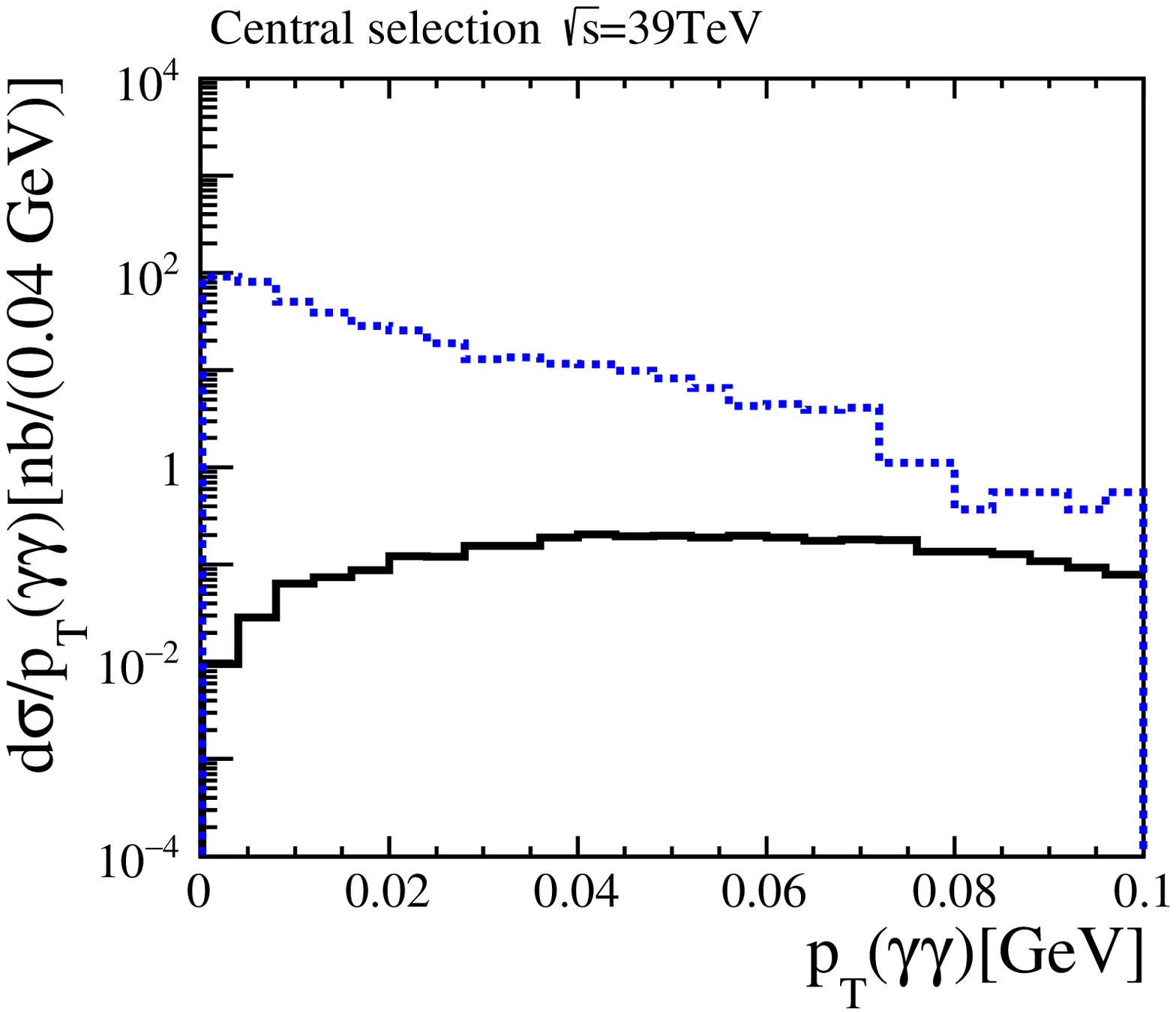}
 \includegraphics[width=0.45\textwidth]{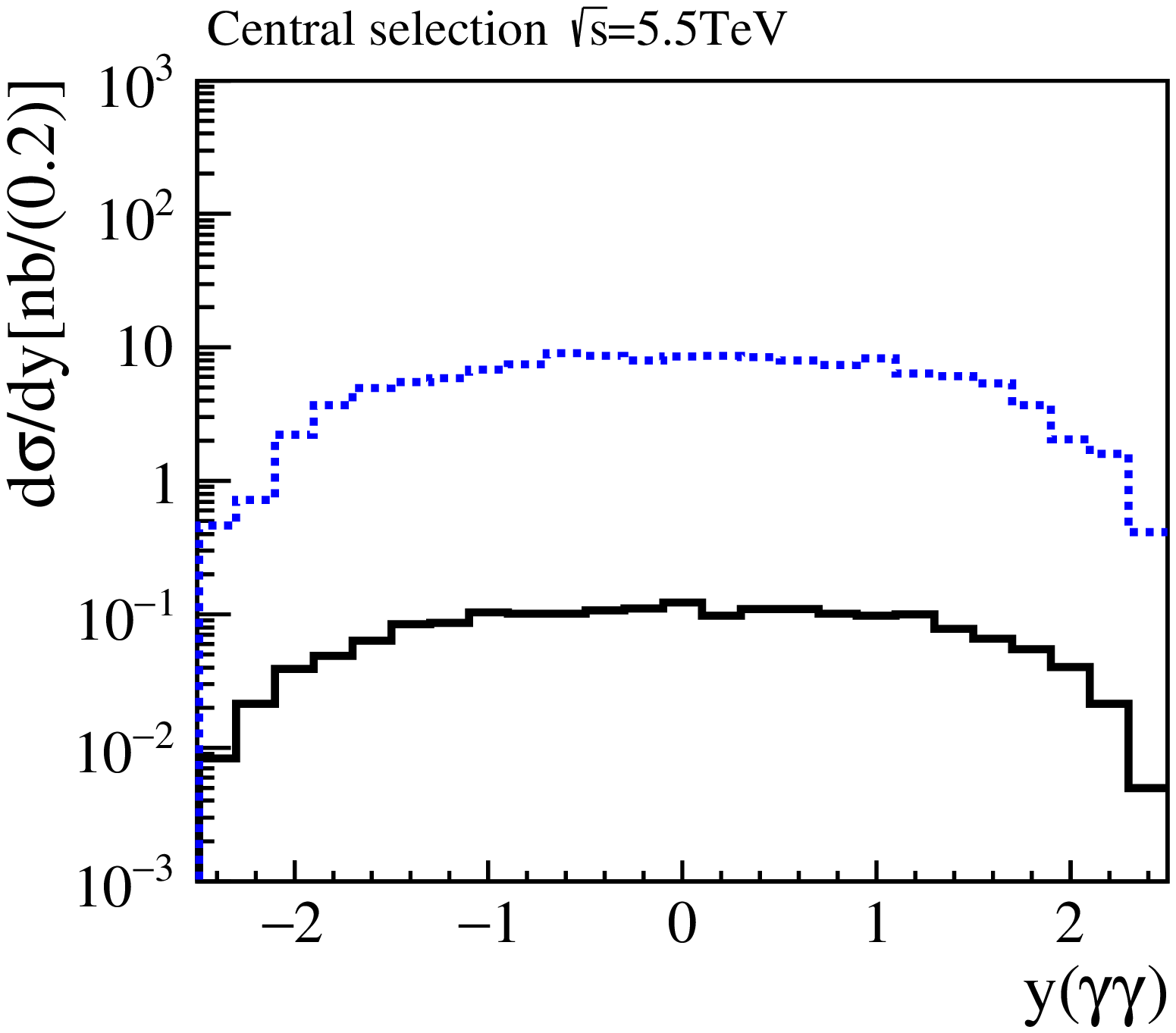}
 \includegraphics[width=0.45\textwidth]{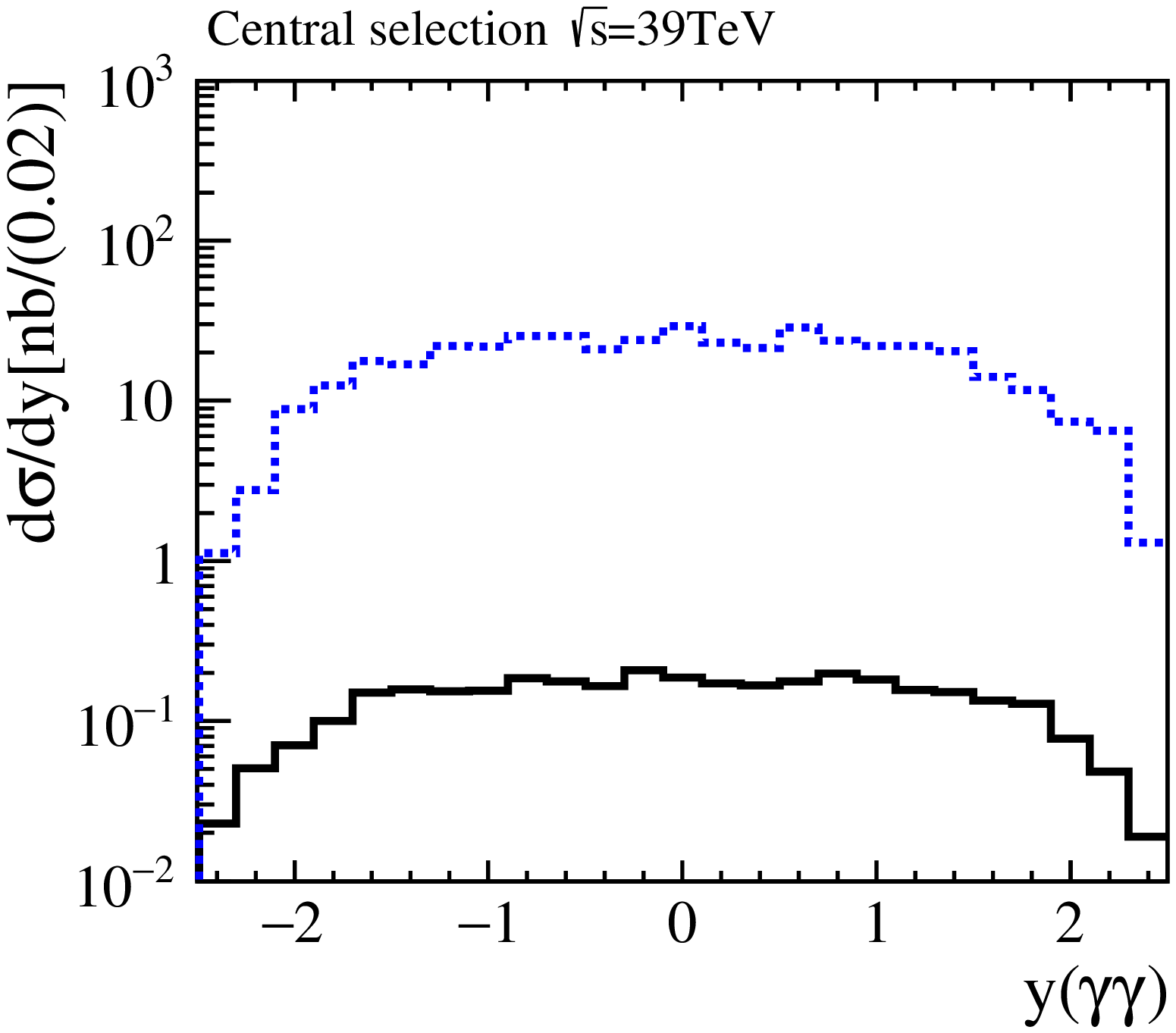}
 \includegraphics[width=0.45\textwidth]{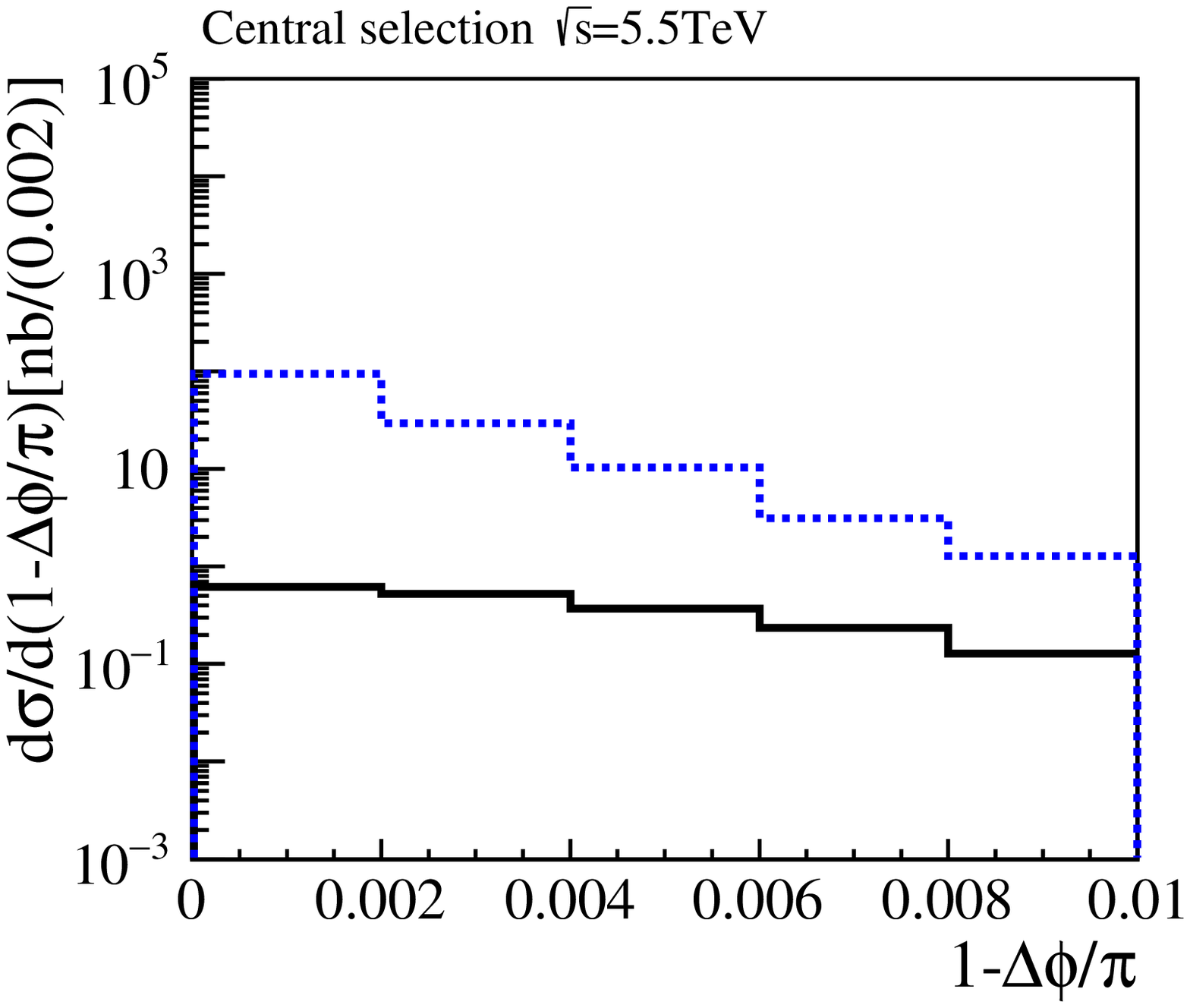}
 \includegraphics[width=0.45\textwidth]{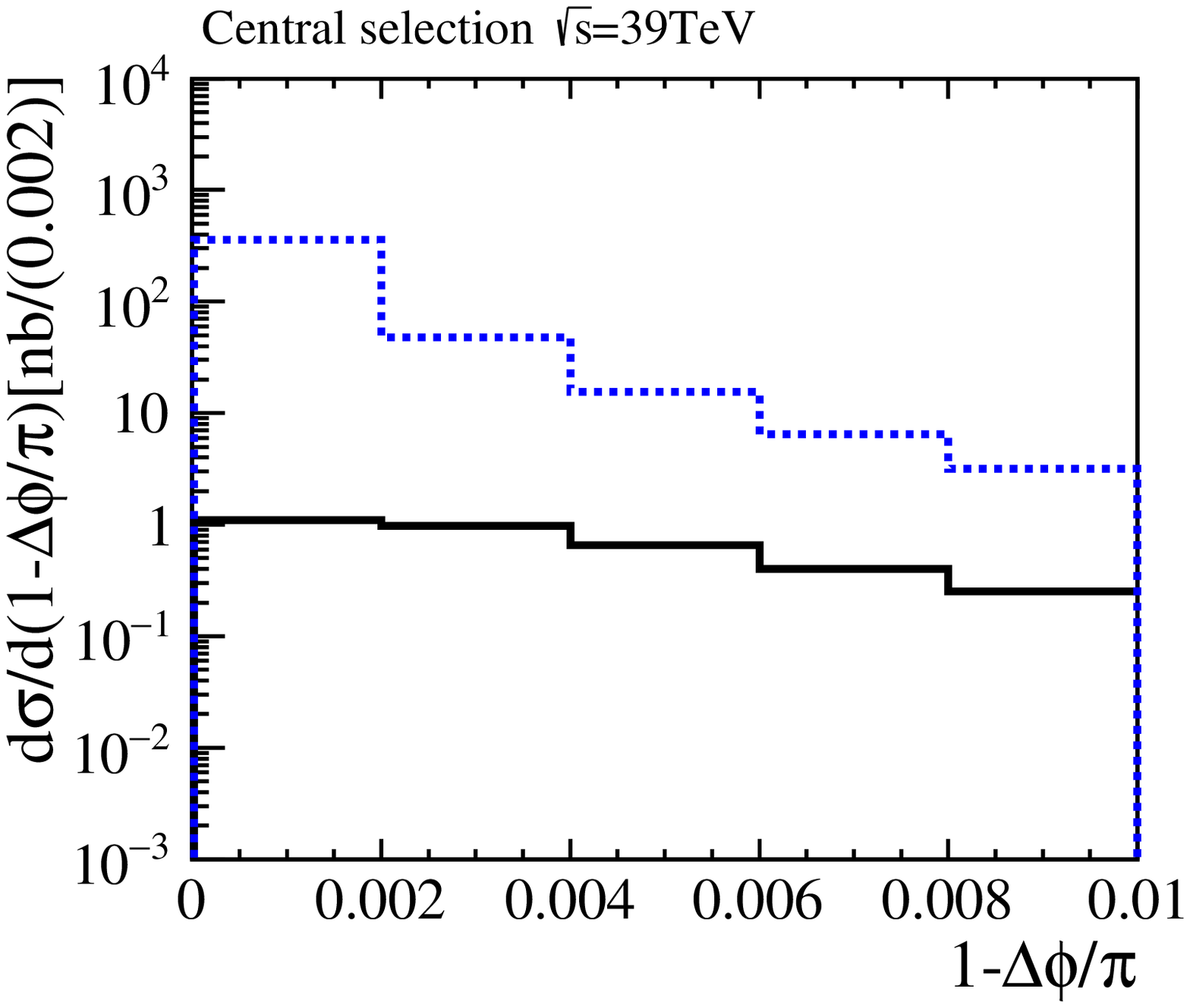}
 \caption{Results for the invariant mass $m_X$, transverse momentum $p_T (\gamma \gamma)$, rapidity $y(\gamma \gamma)$ and acoplanarity distributions considering a central detector and $PbPb$ collisions at the LHC (left panels) and FCC (right panels).}
\label{fig:central}
 \end{figure}
 \end{center} 
 
 \begin{center}
 \begin{figure}[htbp!]
 \includegraphics[width=0.45\textwidth]{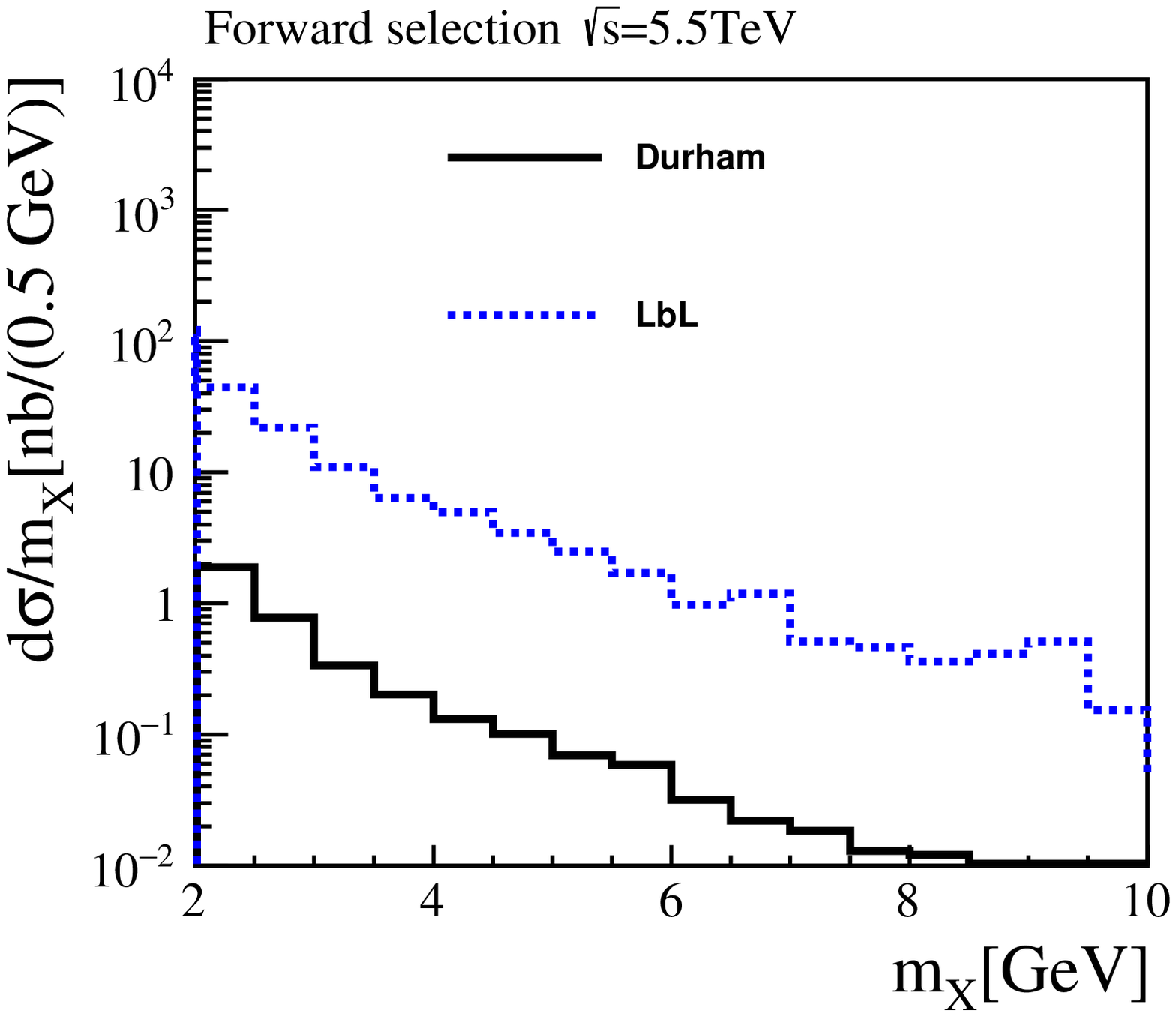}
 \includegraphics[width=0.45\textwidth]{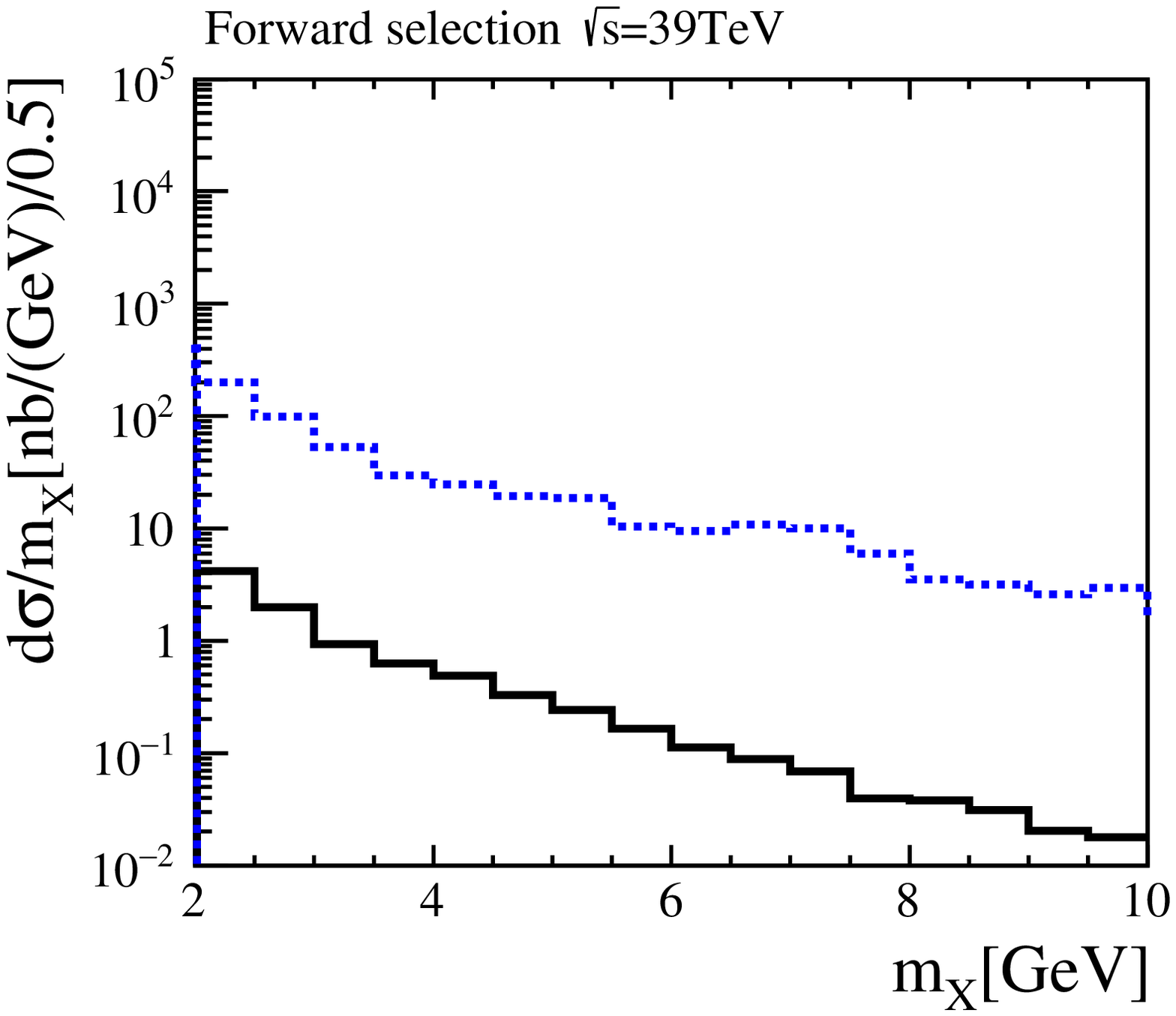}
 \includegraphics[width=0.45\textwidth]{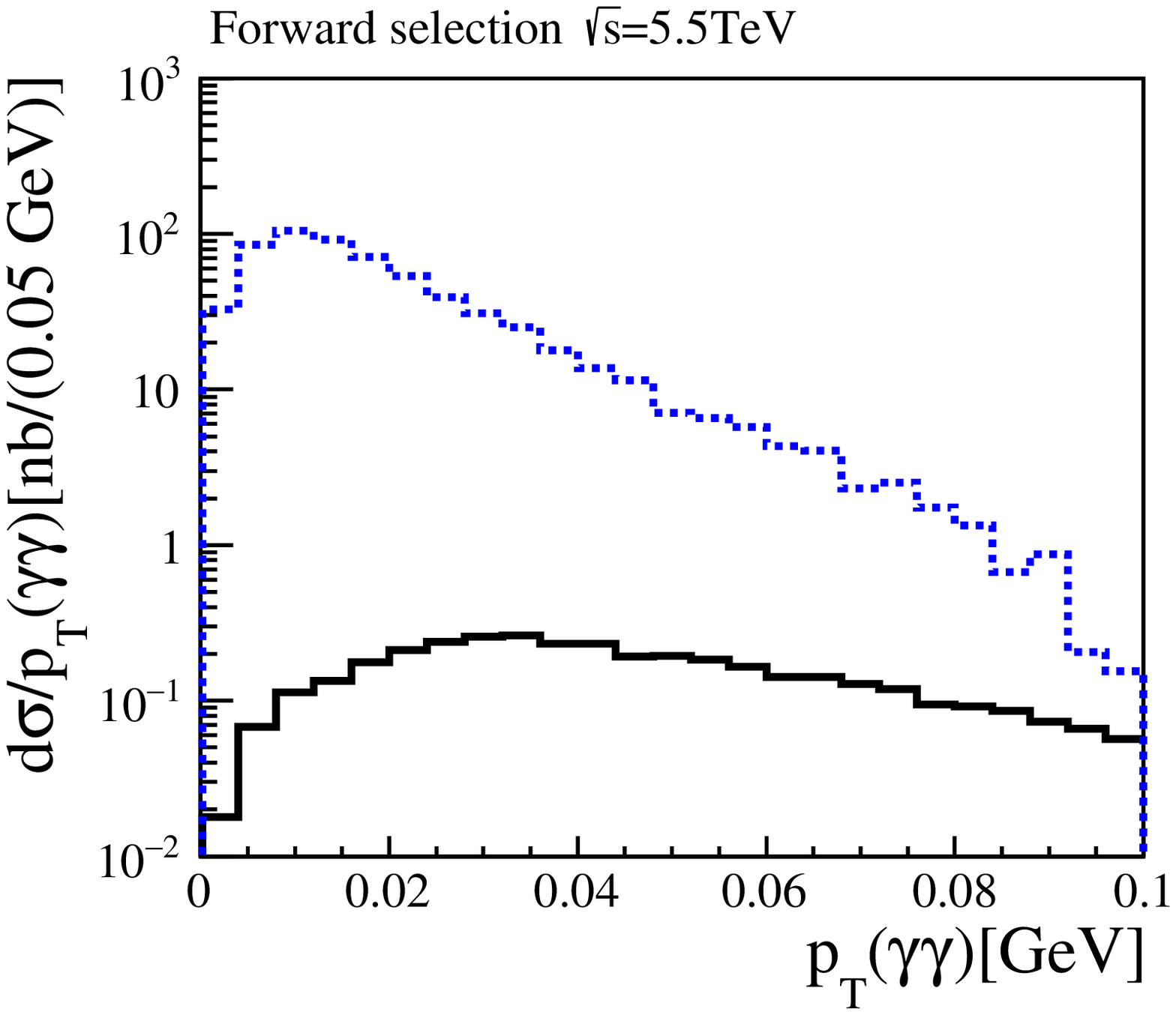}
 \includegraphics[width=0.45\textwidth]{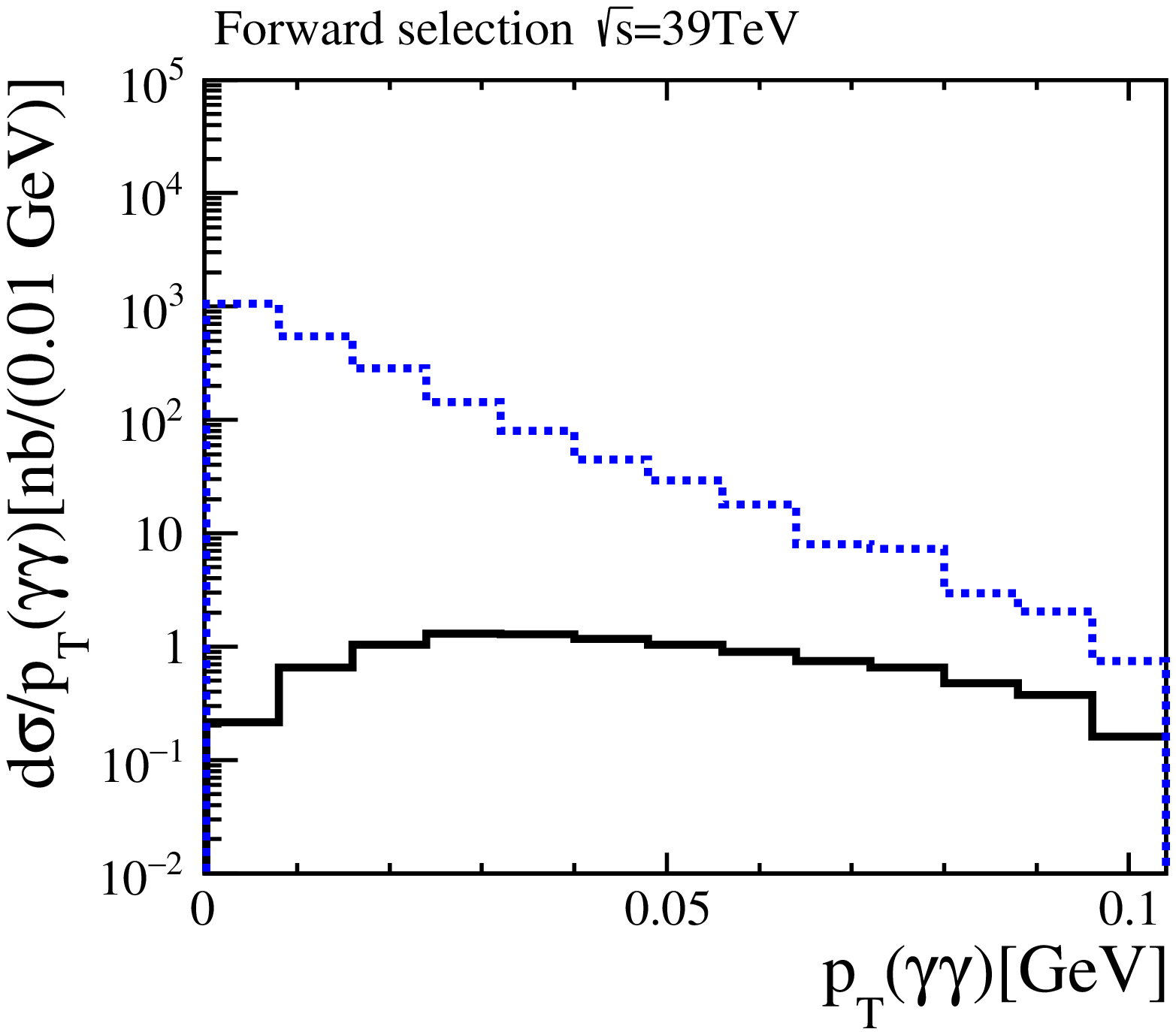}
 \includegraphics[width=0.45\textwidth]{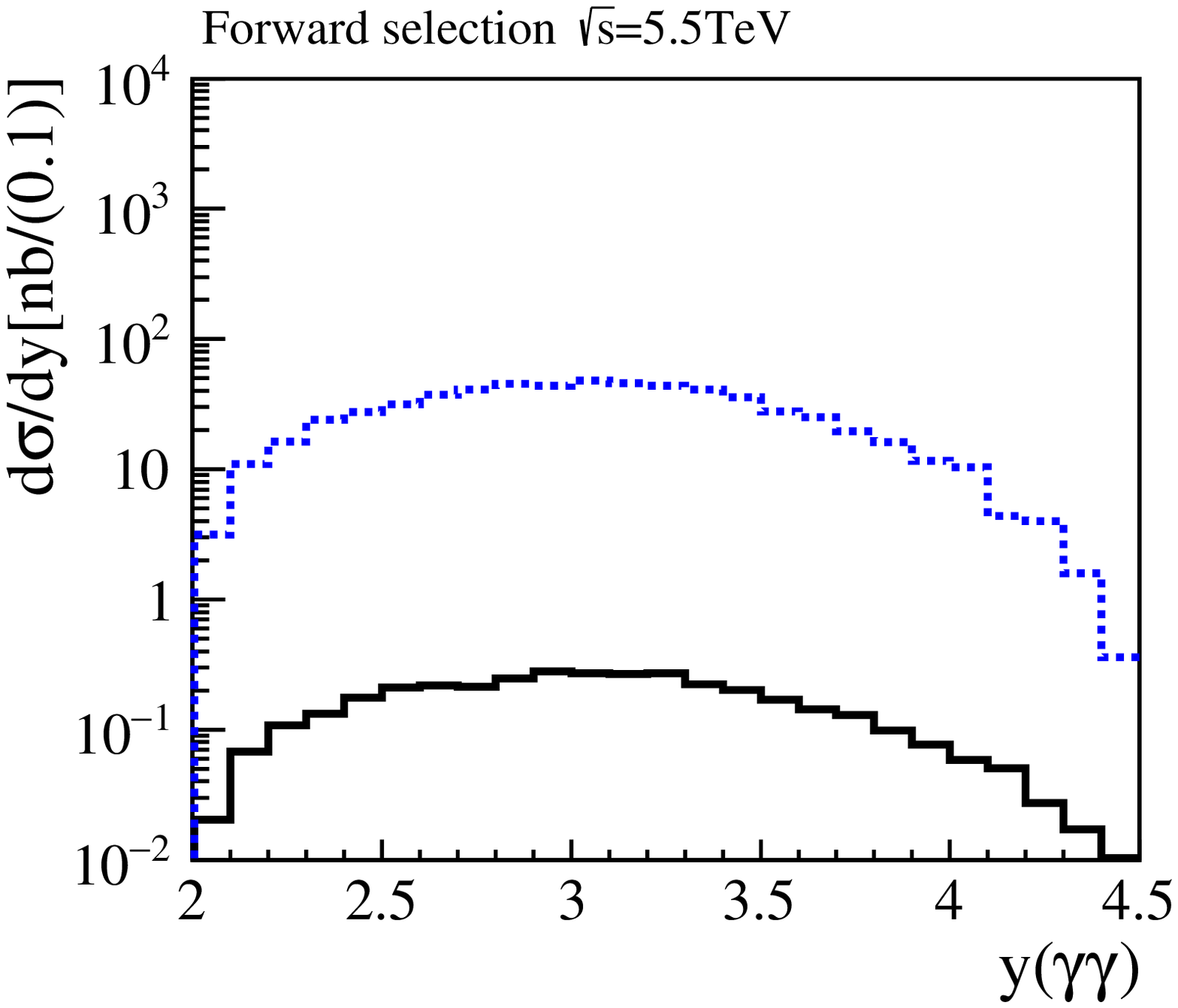}
 \includegraphics[width=0.45\textwidth]{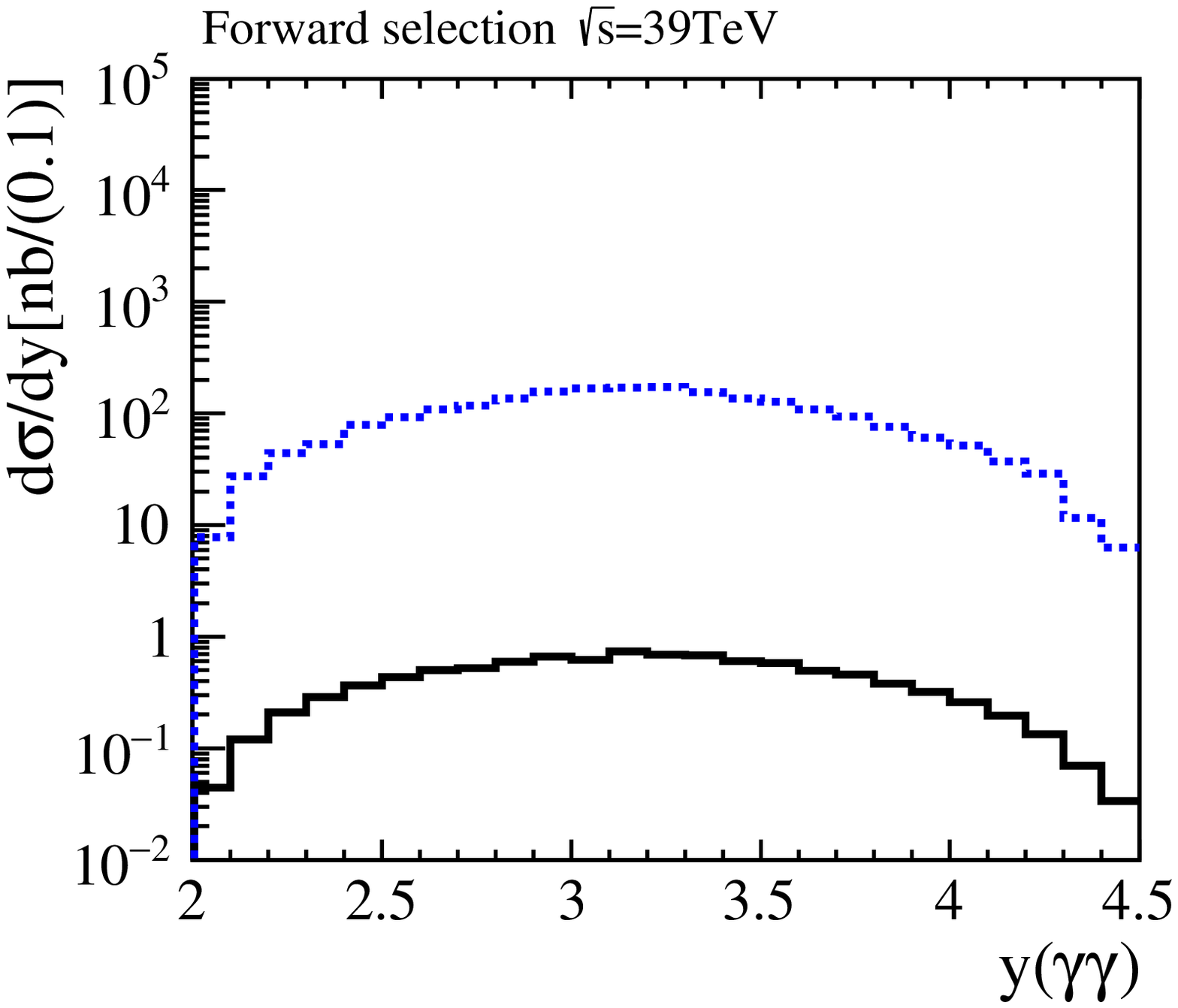}
 \includegraphics[width=0.45\textwidth]{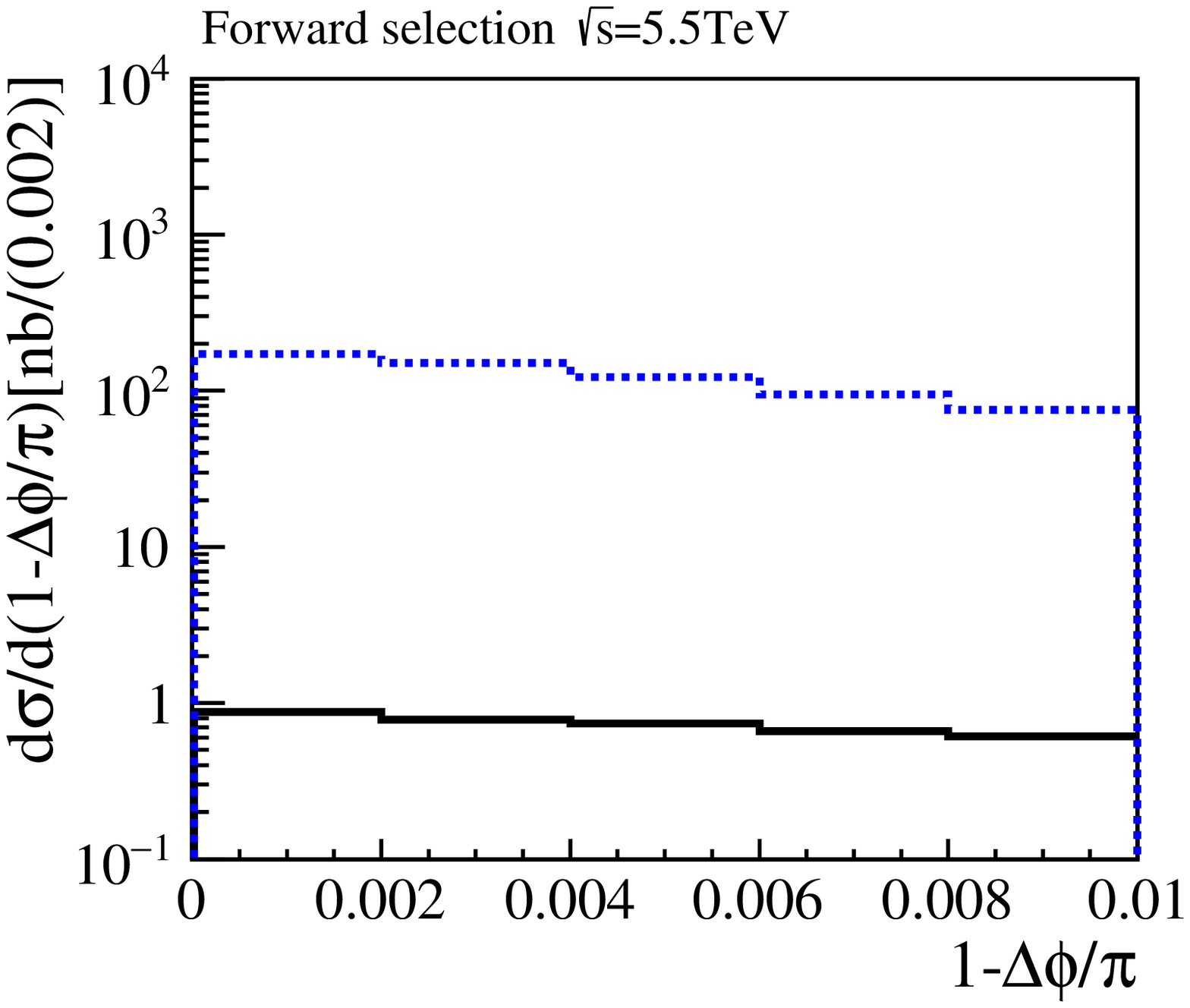}
 \includegraphics[width=0.45\textwidth]{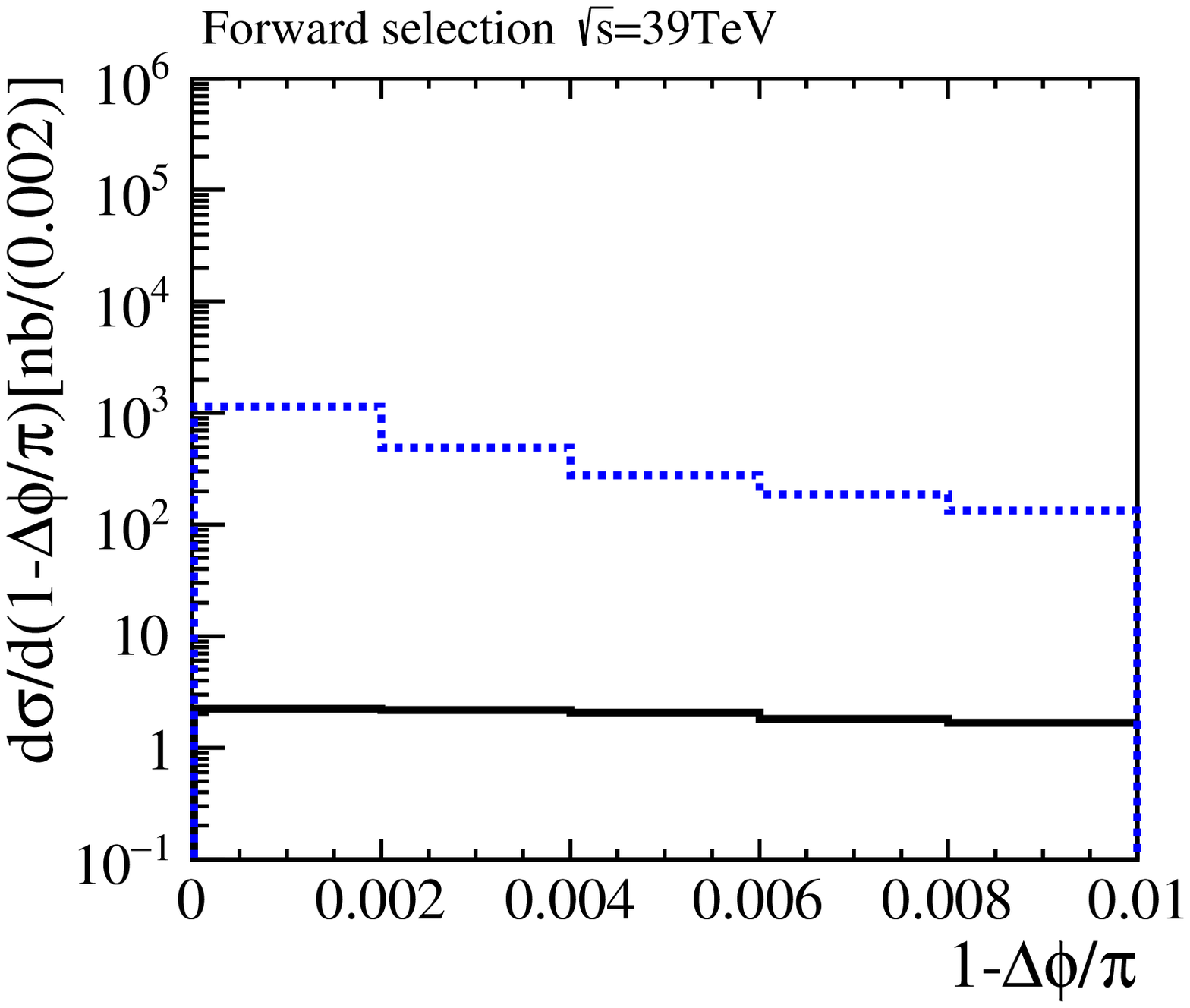}
 \caption{Results for the invariant mass $m_X$, transverse momentum $p_T (\gamma \gamma)$, rapidity $y(\gamma \gamma)$ and acoplanarity distributions considering a forward detector and $PbPb$ collisions at the LHC (left panels) and FCC (right panels).}
\label{fig:forward}
 \end{figure}
 \end{center} 
  
The impact of each of these cuts in the different processes for the LHC, HE -- LHC and FCC energies is presented in the Tables \ref{tab:central} and \ref{tab:forward} for a central and forward detector, respectively. For the central detector configuration, we have that the cut in the invariant mass has a large impact in the LbL and Durham processes. Such impact is smaller in the case of a forward detector, since the events with small invariant masses are included. The LbL and Durham predictions are not strongly modified by the inclusion of the other cuts, unless of the cut on rapidity for a forward detector which suppress the cross section by one order of magnitude. In addition, our results indicate that the inclusion of all cuts fully suppress the contribution of the double diffractive process for the diphoton production. Finally, we predict that the contribution of the Durham model for the exclusive $\gamma \gamma$ production is two orders of magnitude smaller than the LbL process. Such result implies that the LbL process could be studied in the future run of the LHC, as well in the future HE -- LHC and FCC, in a clean environment and reduced background, which will allow a detailed search by Beyond Standard Model physics using this final state.

In Fig. \ref{fig:central}  we present our predictions for the  invariant mass $m_X$, transverse momentum $p_T (\gamma \gamma)$, rapidity $y(\gamma \gamma)$ and acoplanarity distributions considering a central detector and $PbPb$ collisions at the LHC (left panels) and FCC (right panels). We have that the Durham process only becomes competitive for a diphoton system with a large transverse momentum. Similar results, but with a distinct normalization, are obtained for a forward detector, as verified in Fig. \ref{fig:forward}.

 \section{Summary}
 \label{sec:conc}
The high photon -- photon luminosity present in ultraperipheral heavy -- ion collisions become feasible  the experimental analysis of different final state  that can be used to test some of the more important properties of Standard Model (SM) as well to search by BSM physics. One of more interesting final states is the diphoton system, which can be  produced by photon --   and gluon -- induced interactions.  
Although the elementary $\gamma \gamma \rightarrow \gamma \gamma$ and $ g  g \rightarrow \gamma \gamma$ subprocesses have a very tiny cross section, the associated $PbPb$ cross sections become measurable  due to the large number of photons and gluons in the initial state. In this paper we have estimated the contribution of the Light -- by -- Light scattering, Durham and double diffractive processes for the diphoton production. The typical cuts used to select exclusive events were taken into account as well as the acceptance of the LHC detectors. In particular, a detailed analysis of the diphoton production in the kinematical range probed by the LHCb detector was performed by the first time. Moreover, we have presented predictions for the diphoton production in $PbPb$ collisions for the energies of the future High Energy -- LHC and FCC. Our results demonstrated that the contribution of the gluon induced processes can be strongly suppressed by the exclusivity cuts. Consequently, future experimental anaysis of the diphoton production will allow to perform a precise study of the LbL process as well to search by New Physics using this final state.

\begin{acknowledgments}
VPG acknowledge very useful discussions about photon - induced interactions with Gustavo Gil da Silveira, Mariola Klusek-Gawenda and Antoni Szczurek.
This work was  partially financed by the Brazilian funding
agencies CNPq, CAPES,  FAPERGS, FAPERJ and INCT-FNA (processes number 
464898/2014-5 and 88887.461636/2019-00).
\end{acknowledgments}

 \end{document}